\pdfoutput=1
\documentclass[sigconf, 10pt]{acmart}

\AtBeginDocument{%
  }

\graphicspath{{./}{./img/}}
\DeclareGraphicsExtensions{.pdf,.jpeg,.jpg,.png,.eps,.ps}

\usepackage{mathtools}
\usepackage{amsfonts,amsthm}
\usepackage{algorithmic}
\usepackage{array}
\usepackage{mdwmath}
\usepackage{mdwtab}
\usepackage{eqparbox}
\usepackage[font=footnotesize]{subfig}
\usepackage{stfloats}
\usepackage{url}
\usepackage{multirow}

\usepackage[lined, ruled, linesnumbered]{algorithm2e}
\usepackage{enumitem}
\usepackage{pbox}
\usepackage{float}
\usepackage{pgffor}
\usepackage{bm}		%
\usepackage{tabularx}	%
\usepackage{framed}	
\usepackage{pifont}		%
\usepackage{tikz} %
\usepackage{multirow} %
\usepackage{graphicx} %
\usepackage{caption} %
\usepackage{booktabs}

\setcopyright{acmlicensed}
\copyrightyear{2018}
\acmYear{2018}
\acmDOI{XXXXXXX.XXXXXXX}

\acmConference[Conference acronym 'XX]{Make sure to enter the correct
  conference title from your rights confirmation emai}{June 03--05,
  2018}{Woodstock, NY}
\acmISBN{978-1-4503-XXXX-X/18/06}

\newtheorem{definition}{\bf Definition}	%
\newtheorem{observation}{\bf Observation}	%

\newcounter{appdx}
\definecolor{dgreen}{rgb}{0,0.655,0.149}

\newcommand{\redc}[1]{{\color{red}}}

\newcommand{\oursol}{{OrbitChain}}

\definecolor{dgreen}{rgb}{0,0.655,0.149}

\SetKw{Continue}{continue}
\SetKw{Break}{break}

\makeatletter
\newcommand{\nosemic}{\renewcommand{\@endalgocfline}{\relax}}%
\newcommand{\dosemic}{\renewcommand{\@endalgocfline}{\algocf@endline}}%
\let\oldnl\nl%
\newcommand{\nonl}{\renewcommand{\nl}{\let\nl\oldnl}}%
\makeatother

\SetKwInOut{Init}{Initialize}
\SetKwInOut{OptIn}{Optional input}

\setcopyright{none}
\begin{document}

\title{OrbitChain: Orchestrating In-orbit Real-time Analytics of Earth Observation Data}
\author{Zhouyu Li}
\email{zli85@ncsu.edu}
\affiliation{%
  \institution{North Carolina State University}
  \city{Raleigh}
  \country{USA}
}

\author{Zhijin Yang}
\email{zyang44@ncsu.edu}
\affiliation{%
  \institution{North Carolina State University}
  \city{Raleigh}
  \country{USA}
}

\author{Huayue Gu}
\email{hgu2@kennesaw.edu}
\affiliation{%
  \institution{Kennesaw State University}
  \city{Atlanta}
  \country{USA}
}

\author{Xiaojian Wang}
\email{xiaojian.wang@ucdenver.edu}
\affiliation{%
  \institution{University of Colorado Denver}
  \city{Denver}
  \country{USA}
}

\author{Yuchen Liu}
\email{yliu322@ncsu.edu}
\affiliation{%
  \institution{North Carolina State University}
  \city{Raleigh}
  \country{USA}
}

\author{Ruozhou Yu}
\email{ryu5@ncsu.edu}
\affiliation{%
  \institution{North Carolina State University}
  \city{Raleigh}
  \country{USA}
}

\begin{abstract}
Earth observation analytics have the potential to transform many sectors.
However, due to limited ground connections, it currently takes hours to days to download and analyze Earth observation data, diminishing the value of data for time-sensitive applications like disaster monitoring or search-and-rescue.
To enable real-time analytics, we propose {\oursol}, an in-orbit multi-satellite Earth analytics framework.
{\oursol} uses a pipelined design to decompose workflows into analytics functions, and orchestrates constellation-wide resources to finish real-time analytics tasks.
It provides timely insights to Earth sensing applications and enables advanced workflows like in-orbit tip-and-cue.
Hardware-in-the-loop experiments show that {\oursol} can deliver analytics results in minutes, supports up to $60\%$ more analytics workload than existing frameworks, and reduces inter-satellite communication overhead by up to $45\%$.
\end{abstract}



\maketitle

\vspace{-0.15em}
\section{Introduction}
\label{sec:intro}
\noindent
Satellite-based Earth observation has found wide applications such as in urban development and disaster monitoring.
Yet, existing constellations employ satellites merely as passive sensors, capturing data that must be downloaded for ground analytics.
This paradigm causes significant delay and data loss due to limited ground-satellite connection windows and bandwidth.
These delays range from
several hours to days~\cite{pbc2018planet}, making the data unusable by time-sensitive tasks like maritime surveillance~\cite{eoapplication} and disaster monitoring~\cite{leyva-mayorga2023satellite}.
Furthermore, most data collected by a satellite will eventually be dropped due to bufferbloat~\cite{wang2024infiltrating}, limiting coverage and data availability within a timeframe~\cite{denby2023kodan}.

Recent advances in nanosatellites have enabled on-board computation on Earth observation satellites~\cite{donahue2024planet}, which gave rise to the concept of orbital edge computing (OEC)~\cite{denby2019orbital}.
OEC alleviates downlink channel constraints by filtering low-value data such as cloud-coverage images~\cite{denby2023kodan}, or compressing data before downlink~\cite{sun2025deepspace}.
However, major analytics are still ground-based with downloaded raw data, resulting in long analytics delay that hinders time-sensitive applications.%

Fully in-orbit analytics could ideally achieve the lowest end-to-end latency for Earth observation applications, but are limited by per-satellite computation and energy budgets.
Existing work has explored parallelism to balance analytics workloads across multiple satellites~\cite{denby2020orbital}, where each satellite processes a part of the ground region, and results are aggregated on the ground.
However, as we show in Section~\ref{sec:preliminary_study:oec}, such independent parallelism inefficiently utilizes in-orbit resources, and cannot support complex analytics workflows. %

We develop \textbf{\oursol}, a multi-satellite OEC framework for in-orbit real-time Earth observation analytics.
{\oursol} utilizes a chain of satellites with overlapping ground coverage~\cite{denby2019orbital,denby2020orbital,cheng2024eagleeye}, and orchestrates sensing and analytics pipelines utilizing both data and computation parallelism.
Analytics workflows are decomposed into analytics functions, which are distributed across satellites and coordinated via inter-satellite links to enable efficient resource utilization.
Based on offline profiling, {\oursol} employs an optimization engine that determines function deployment and resource allocation, and proactively routes analytics workloads to minimize inter-satellite communications.
We implemented and evaluated {\oursol} on prevalent satellite on-board systems (NVIDIA Jetsons and Raspberry Pis)
~\cite{ltdraspberry,2020lockheed}.
Results show that {\oursol} can complete up to $60\%$ more analytics workloads than existing frameworks, and on average saves up to $45\%$ inter-satellite communication overhead.

Our main contributions are summarized as follows:
\vspace{-0.3em}
\begin{itemize}
    \item We design {\oursol}, a multi-satellite OEC framework that enables fully in-orbit real-time sensing and analytics pipelines for complex Earth analytics tasks.
    \item We develop a graph-based workflow orchestrator that profiles and deploys analytics functions across multiple satellites, and orchestrates cross-satellite resources and workload routing to achieve high analytics throughput with minimal communication overhead.%
    \item We evaluate {\oursol} on an OEC testbed with hardware used in space, and demonstrate its superior performance over existing OEC frameworks.
\end{itemize}
\vspace{-0.3em}

\section{Background}
\label{sec:background}
\subsection{Earth Observation Analytics Tasks}
\begin{figure}[t]
\centering
\includegraphics[width=0.45\textwidth]{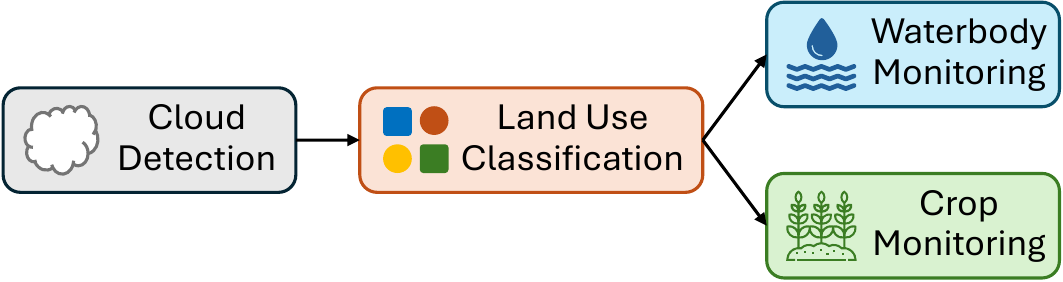}
\vspace{-1em}
\caption{The models and data flow in an Earth observation workflow for farmland flood monitoring.}
\label{fig:application}
\end{figure}
Growing demand for Earth analytics has driven rapid growth in Earth observation constellations~\cite{dove,rapideye,2016sentinel2}.
In Earth observation, low-Earth-orbit (LEO) satellites act as sensors that capture and downlink remote sensing data for ground-based analytics.
Driven by projected economic impacts, commercial providers such as Planet Labs~\cite{pbc2018planet} now offer custom analytics requested by users.
For a user request, related Earth sensing data is downlinked during satellite-ground connections and processed through analytics functions such as cloud filtering, context classification and target object detection~\cite{tao2024known}, usually implemented with deep neural networks~\cite{denby2023kodan,du2025earth,yangefficient}.
Fig.~\ref{fig:application} shows an example farmland flood monitoring workflow that first filters cloud coverage, then identifies waterbody and farmlands, and finally analyzes flood and crop conditions.

The biggest issue with current Earth observation constellations is the long delay for data downloading, due to restricted downlink bandwidth and scarce connection windows~\cite{denby2023kodan, sun2025spacesched}.
For instance, a Sentinel-2 satellite generates about 2.7~TB data per day, yet its five ground stations can only download 1 TB of in-orbit data daily~\cite{2016sentinel2}. 
The remaining data are buffered, leading to up to 30 days of delay for data downloading and rendering them unusable for time-sensitive tasks.

\subsection{Orbital Edge Devices}
\label{sec:background:oec_device}
Advances in edge computing have enabled devices like Raspberry Pis and NVIDIA Jetsons to serve as on-board computing units on recent satellites~\cite{ltdraspberry,2020lockheed}. 
Recent research has proposed using these devices to host various analytics models for on-board Earth observation analytics.
For example, Kodan~\cite{denby2023kodan} performs data filtering in orbit to save downlink bandwidth by downloading only high-value data such as cloud-free images.

Serval~\cite{tao2024known} divides the Earth observation analytics workload between in-orbit and ground execution based on the type of query (whether it is static or dynamic). %
Earth+~\cite{du2025earth} performs in-orbit data compression based on the difference between historical and new sensing data.
DeepSpace~\cite{sun2025deepspace} compresses images on the satellite and uses super-resolution to recover downloaded low-resolution images.

However, most existing work uses OEC mainly for data preprocessing and still runs complex analytics on the ground, which leads to substantial end-to-end delays.

\subsection{Inter-satellite Links}
\label{sec:background:inter_sat_links}
\noindent
Inter-satellite links enable cross-satellite coordination and data sharing without ground assistance.
In most constellations, satellites establish \emph{space relay} links and connect only two their nearest neighbors~\cite{intersatellite}.
Supported data rates range from several Kbps to several Gbps~\cite{sodnik2010optical}. 
For example, LoRa radios used on many LEO satellites typically provide 5 Kbps to 50 Kbps of data rate~\cite{gadre2022lowlatency,mehdi2023lorawan}, while S-band links can reach up to 2 Mbps~\cite{aac_pulsar_stx}.
One big limiting factor for inter-satellite communications is energy consumption.
A recent benchmark reports up to $18$W while transmitting and near zero power in idle mode~\cite{xing2024deciphering}. 
The use of inter-satellite links should therefore be carefully planned and minimized.

\section{In-orbit Earth Observation Analytics: Limitations and Opportunities}
\label{sec:preliminary_study}
Earth observation analytics frameworks fall into two categories, ground-assisted and fully in-orbit, depending on whether the ground station or cloud participates.
Ground-assisted frameworks fail to deliver real-time results due to scarce contact windows to the ground~\cite{denby2023kodan}. 
Appendix~\ref{appdx:ground_assist_framework} provides a detailed discussion with supporting data.
This motivates us to focus on fully \emph{in-orbit Earth observation analytics} to deliver real-time insights.

\begin{figure}[t]
\centering
\subfloat[Ground track frame \& inter-frame time (frame deadline) $\Delta_f$.]{\includegraphics[width=0.232\textwidth]{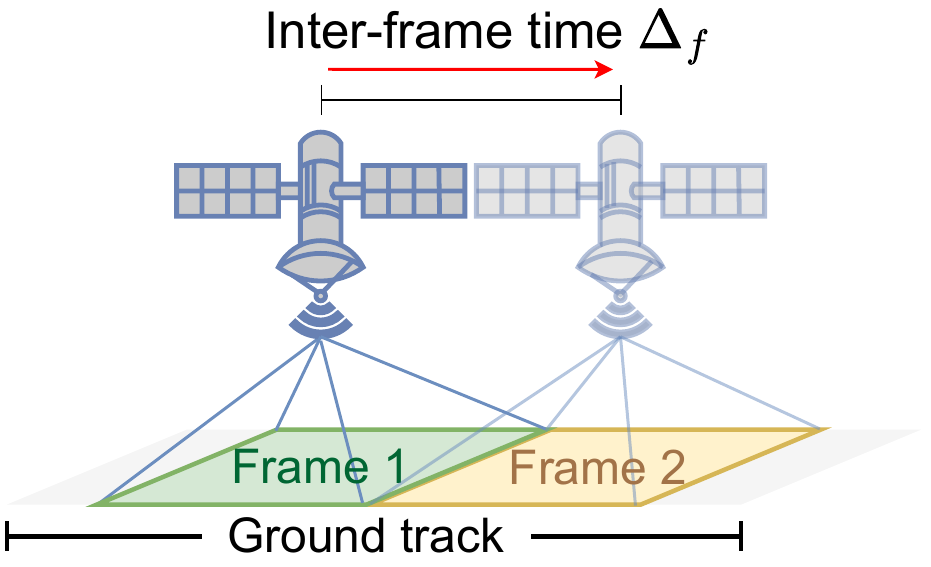}%
\label{fig:moving_mode:frame}}
\hfil
\subfloat[The leader-follower constellation design.]{\includegraphics[width=0.215\textwidth]{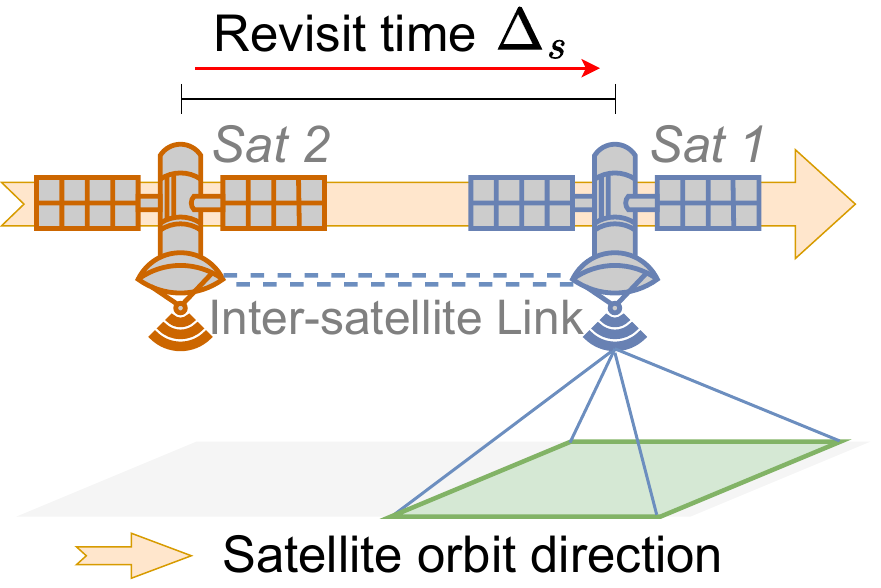}%
\label{fig:moving_mode:leader_follower}}
\vspace{-1em}
\caption{(a) Contiguous ground track frames taken by \emph{one} satellite. (b) Adjacent satellites in the same orbit covering \emph{overlapping} areas along their ground tracks.}
\label{fig:moving_mode}
\end{figure}

\subsection{OEC Preliminaries}
\label{sec:oec_prelim}

As shown in Fig.~\ref{fig:moving_mode}\subref{fig:moving_mode:frame}, an OEC satellite uses its sensor to continuously capture data of the ground surface along its \emph{ground track}. 
The area covered by one capture is called a ground track \emph{frame}.
The inter-frame time $\Delta_f$ represents the minimum time between two non-overlapping captures from the same satellite.
After capturing a frame, the satellite forwards it to its onboard compute unit for analysis.
To prevent buffer buildup, analysis must finish before the next capture. 
Thus the inter-frame time $\Delta_f$ is also called the \emph{frame deadline}.

Due to fast satellite movement and limited on-board computing resources, a single OEC satellite usually cannot analyze a full frame within the frame deadline.
A common OEC practice is dividing a frame into small \emph{tiles}, each being analyzed independently~\cite{denby2020orbital,denby2023kodan}.
This enables parallel processing within a satellite and across collaborating satellites.
In multi-stage workflows like Fig.~\ref{fig:application}, tiling also enables early dropping of low-value tiles, such as those covered by clouds.

Multi-satellite OEC enables in-orbit Earth observation analytics by distributing workload across a constellation~\cite{denby2020orbital}.
The leader-follower constellation design is widely used in Earth observation constellations~\cite{denby2020orbital,cheng2024eagleeye}.
In this design, $N_s$ satellites are evenly spaced along an orbit or orbit segment, with revisit time $\Delta_s$ between consecutive satellites over the same ground track location.
Each satellite captures the same or largely overlapping ground track frames in a sequence~\cite{denby2020orbital}.
The satellites in a constellation can thus share the workload of frame processing, as we discuss next.

\subsection{Multi-satellite OEC Limitations}
\label{sec:preliminary_study:oec}

\noindent
Existing multi-satellite OEC frameworks can be divided into two categories: \emph{data parallelism}, and \emph{compute parallelism}.

\begin{figure}[t]
\centering
\subfloat[Data parallelism: tile load balancing across satellites.]
{\includegraphics[width=0.2\textwidth]{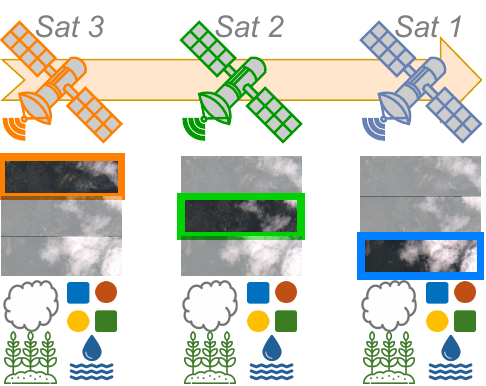}
\label{fig:data_parallel:visualization}}
\hfil
\subfloat[Model inference latency when co-hosted with other models.]
{\includegraphics[width=0.23\textwidth]{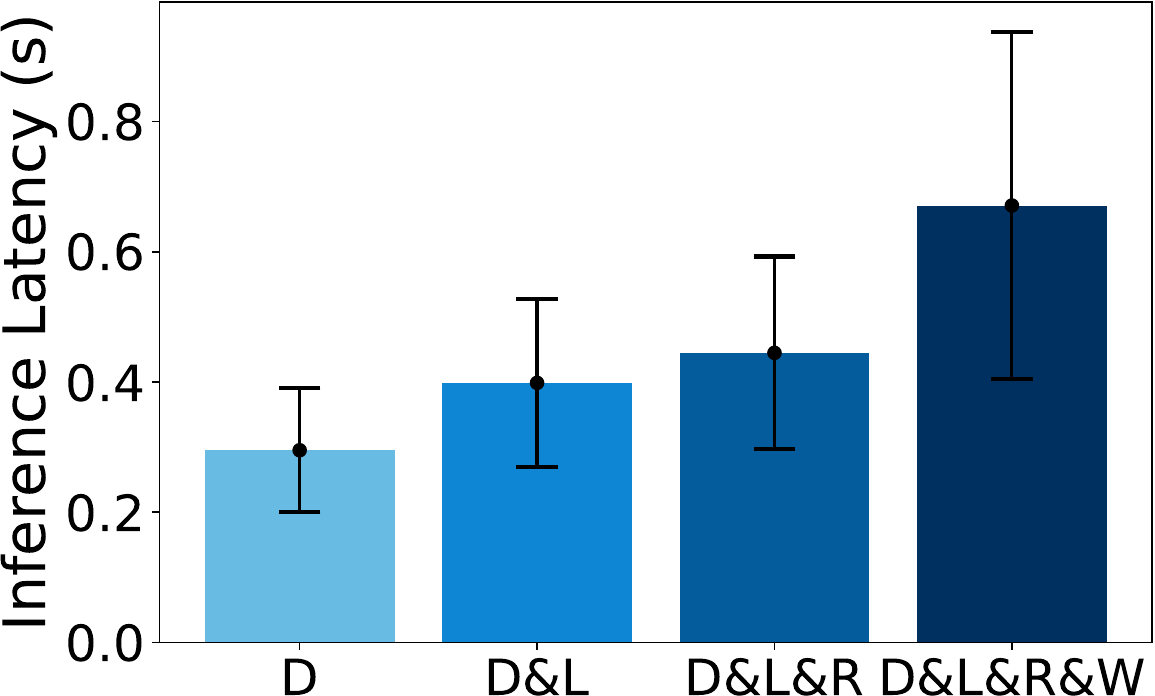}
\label{fig:data_parallel:resource_contention}}
\hfil
\vspace{-1em}
\caption{(a) Data parallelism: each satellite hosts all models and processes different tiles. (b) Cloud detection latency, when co-hosted with other models on the same satellite. Error bars show standard deviation over 10 runs. (D: cloud detection; L: land use classification; R: crop monitoring; W: water monitoring.)}
\label{fig:data_parallel}
\end{figure}
\noindent

\textbf{Data parallelism.}
In data parallelism, each satellite processes a subset of tiles from one or multiple frame, and the constellation collectively processes all tiles~\cite{denby2020orbital}.
This requires no inter-satellite communication because workload assignment can be pre-defined.
Fig.~\ref{fig:data_parallel}\subref{fig:data_parallel:visualization} shows an example with three satellites that evenly split the workload for one frame.
\noindent\textbf{\emph{Limitations:}}
A key limitation of data parallelism is that every satellite runs the full set of analytics models.
As the workflow grows in complexity, co-located models increase resource contention, and can prevent complex workflows from being implemented.
In Fig.~\ref{fig:data_parallel}\subref{fig:data_parallel:resource_contention}, we deploy YOLOv8n-based models~\cite{ultralytics,liu2024inorbit} from Fig.~\ref{fig:application} on a NVIDIA Jetson unit.
We observe substantial slowdowns in model inference time when co-located with other models.
Especially, when combined memory consumption exceeds device capacity, the workflow cannot be instantiated, as we also observe in Section~\ref{sec:evaluation}.

\begin{figure}[t]
\centering
\subfloat[Compute parallelism: each satellite hosts part of a pipeline.]
{\includegraphics[width=0.215\textwidth]{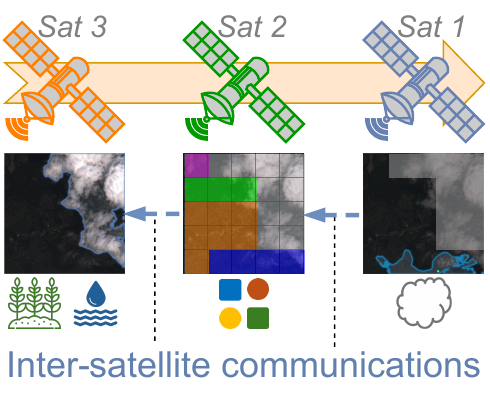}
\label{fig:computational_parallel:visualization}}
\hfil
\subfloat[Time for different models to analyze $100$ $640\text{px} \times 640\text{px}$ images.]{\includegraphics[width=0.225\textwidth]{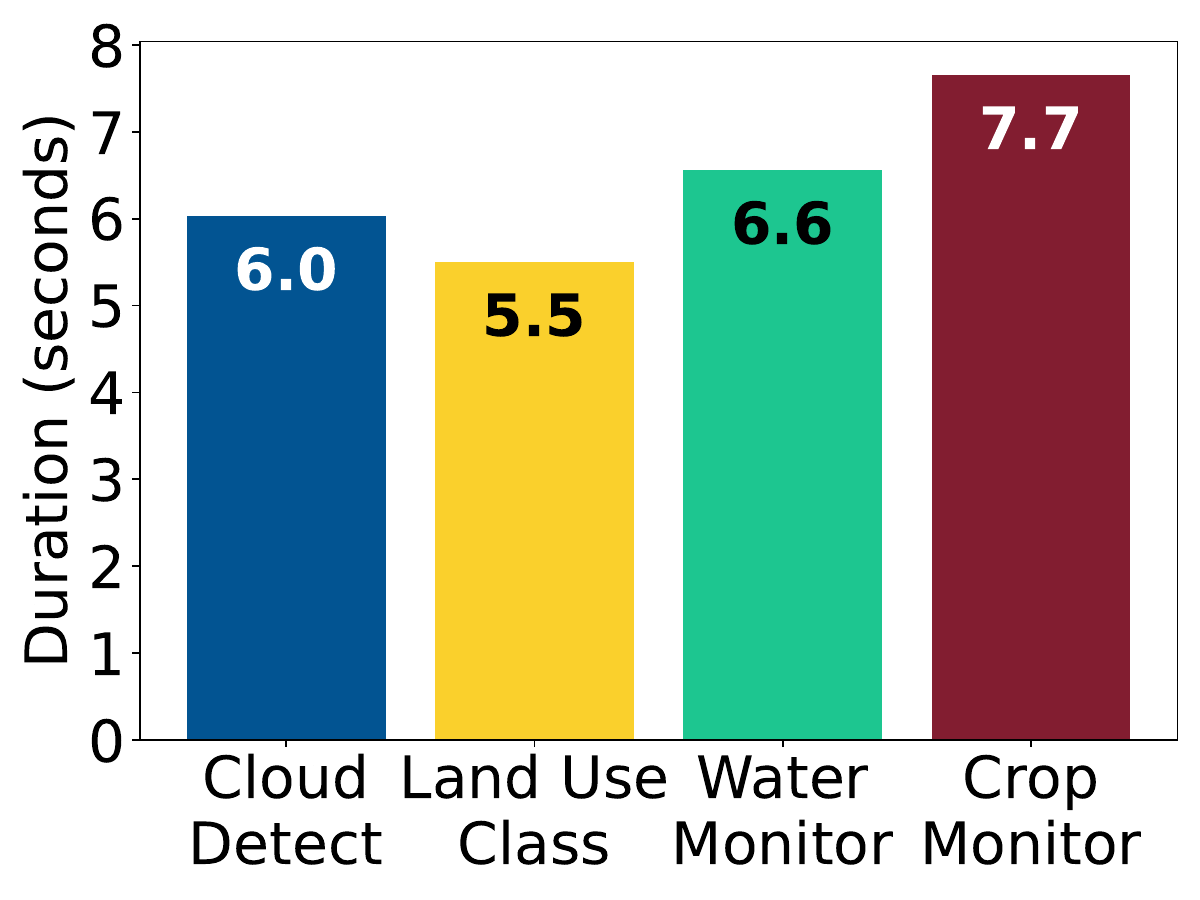}
\label{fig:computational_parallel:speed_difference}}
\vspace{-1em}
\caption{(a) Compute parallelism: models hosted on different satellites use inter-satellite communications to enable pipeline processing. (b) Heterogeneous processing times yield different throughput across models.}
\label{fig:computational_parallel}
\end{figure}

\textbf{Compute parallelism.}
To host complex workflows, compute parallelism places different analytics models on different satellites~\cite{valente2023optimal}.
Fig.~\ref{fig:computational_parallel}\subref{fig:computational_parallel:visualization} shows an example where four analytics models are placed across three satellites.
Inter-satellite communications are required to transmit raw and intermediate data among models hosted on different satellites.
\textbf{\emph{Limitations:}}
Naive compute parallelism faces two challenges.
First, most analytics models require the raw sensing data at least as part of their inputs, which forces transmitting raw data tiles between satellites.
As we show in Section~\ref{sec:system:profiling}, raw data transmission incurs huge inter-satellite communication overhead, leading to excessive bandwidth usage and energy consumption.
Second, models also have varying processing speeds, as shown in Fig.~\ref{fig:computational_parallel}\subref{fig:computational_parallel:speed_difference}.
This creates bottlenecks for processing throughput when some models saturate their capacities while others remain underutilized.
As a result, workflow throughput is limited by the slowest model.
A sluggish model may cause the entire workflow to miss the frame deadline and lead to buffer overflow.

\subsection{Opportunities and Challenges}
\label{sec:preliminary:oppo_and_challenge}

\noindent
We identify several opportunities for realizing in-orbit real-time analytics.
One opportunity is to combine the strengths of data and compute parallelism, balancing between flexible analytics function placement and communication overhead.
The key insight is to exploit the overlapping views of adjacent satellites to drastically reduce communication overhead while allowing efficient multi-satellite collaborative analytics.
Second, resource isolation techniques such as \texttt{cgroups} and containerization can improve performance predictability with explicit resource allocation~\cite{linux_cgroup_memory_doc,kubernetes_resource_management,docker_resource_constraints}.
Leveraging predictable satellite orbits, proactive workload and performance profiling and intelligent planning, it is possible to achieve seamless, automated cross-satellite collaboration via offline orchestration, without extensive in-orbit coordination.
\textbf{Challenges.}
The key to achieving efficient in-orbit analytics is careful \emph{offline planning and orchestration}.
To enable performance-oriented analytics, one key requirement is to finish data processing as they are continuously generated, while avoiding bufferbloat.
This requires accurate modeling of model performance and workload, along with careful planning of resource allocation and workload distribution to avoid resource contention and congestion.
The orchestration must respect limited onboard computation resources, storage, and energy budget.
Cross-satellite communications should be minimized when routing workloads to conserve on-board energy.
Offline planning must also account for changes in satellite orbits and differences in each satellite's captured view when deciding model placement and workload distribution.
As this leads to a convoluted set of mixed-integer decisions and constraints, the underlying planning problem is firmly within the NP-hard category, warranting a combination of novel design and system implementation.

\section{Modeling Analytics Workloads}
\label{sec:framework0}

\noindent
In this section, we first introduce the abstraction of Earth observation analytics workflows and the satellite sensing and analytics pipelines. 
An Earth observation workflow is decomposed into \emph{analytics functions}, each performing a specific task using pre-trained models.
These functions can then be deployed and orchestrated into various sensing and analytics pipelines, implemented using constrained on-board resources and based on offline performance profiling data.
In Section~\ref{sec:framework}, we utilize these abstractions to develop the in-orbit real-time Earth observation analytics framework.

\begin{figure}[t]
\centering
\includegraphics[width=0.25\textwidth]{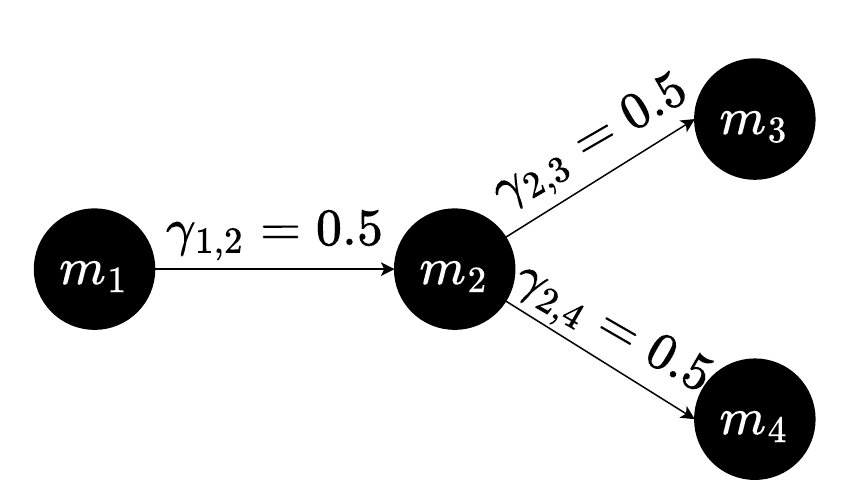}
\vspace{-1em}
\caption{Workflow graph and distribution ratios.}
\label{fig:abstraction}
\end{figure}

\subsection{Decomposing Analytics Workflows}
\label{sec:system:application_abstraction}

An Earth observation analytics workflow consists of multiple image/data analytics models.
In this work, we abstract each model and its additional data pre- or post-processing operations as an \emph{analytics function}. 
A simple Earth observation analytics workflow model employed in the literature is an analytics function chain~\cite{tao2024known}.
We adopt the more expressive directed acyclic graph (DAG)-based model that is more commonly adopted in other analytics workflows, taking into account functions that can process the same or different data in parallel.
Fig.~\ref{fig:application} shows such an example where the same type of land (farm field) can be analyzed independently by a crop monitoring model and a waterbody monitoring model, to jointly decide the impact of flood on crops.

\begin{definition}[Workflow graph]
    An workflow graph $G_A = \left(\mathbf{M}, \mathbf{E}\right)$ has each node $m_i \in \mathbf{M}$ representing an analytics function. A directed edge $e_{i,i'} = \left(m_i, m_{i'}\right) \in \mathbf{E}$ indicates that $m_{i'}$ is a downstream task of $m_i$ and uses $m_i$'s intermediate results for its own processing.

\end{definition}

Fig.~\ref{fig:abstraction} decomposes the analytics workflow from Fig.~\ref{fig:application} into four analytics functions: cloud detection ($m_1$), land use classification ($m_2$), waterbody monitoring ($m_3$), and crop monitoring ($m_4$). 
The data flows in the workflow are represented by directed edges $\left(m_1, m_2\right)$, $\left(m_2, m_3\right)$, and $\left(m_2, m_4\right)$.%

\textbf{Modeling analytics workload.}
As introduced in Section~\ref{sec:preliminary_study}, Earth observation data are tiled before being analyzed.
The processing workload of an analytics function is thus proportional to the number of tiles that it must process.
As in Fig.~\ref{fig:abstraction}, we use the \emph{distribution ratio} on each directed edge $(m_i, m_{i'})$ to represent the approximate amount of analytics workload that $m_i$ brings to $m_{i'}$, which is measured as \emph{the average number of tiles that $m_i$ outputs to $m_{i'}$ per input tile of $m_i$}.
For instance, the cloud detection function ($m_1$) may on-average drop half of the tiles covered by clouds. 
The land use classification function ($m_2$) could then identify only $50\%$ of the remaining filtered tiles as farmland areas and only send these to the waterbody monitoring ($m_3$) and crop monitoring ($m_4$) functions. 
As we show later, these ratios will help us model the end-to-end workload on every analytics function in an end-to-end pipeline per input tile.

\textbf{\textit{Remark:}}
While the distribution ratio of each edge is generally unknown for any specific frame, their statistical values can be estimated by profiling with historical data.
With sufficient on-board buffer, temporary deviation from the statistical estimates can be absorbed without causing long-term buffer build-up.
For unknown locations, the most conservative approach is to set all distribution ratios initially to one, ensuring the full input traffic can be handled. 
As runtime data accumulate, these ratios can be adaptively estimated or predicted by an advanced machine learning model. 
Here we do not explore the distribution ratio in further depth, leaving it as an opportunity for future research.

\begin{figure}[t]
\centering
\includegraphics[width=0.48\textwidth]{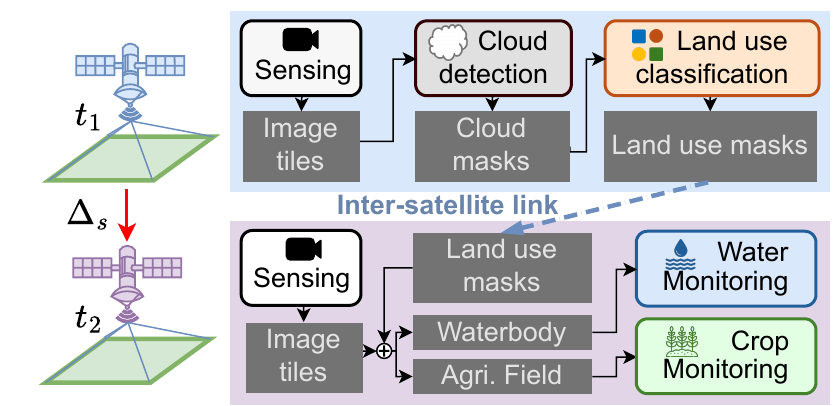}
\vspace{-1.5em}
\caption{Example of a sensing and analytics pipeline involving two satellites, where the inter-satellite link is used to carry inter-mediate results but not raw data.}
\label{fig:orbitchain:sensing_and_analytics}
\vspace{-0.5em}
\end{figure}

\subsection{Sensing and Analytics Pipeline}
\label{sec:sense_ana_pipe}

A core feature of the leader-follower constellation design is the overlapping views across a sequence of satellites along the same orbit, which enables the \emph{data parallelism} as in Section~\ref{sec:preliminary_study}.
In our framework, we utilize this feature to minimize the amount of inter-satellite communications for multi-satellite collaborative analytics.
Specifically, each satellite is pre-deployed with a \emph{sensing function}, which captures, pre-processes, and tiles Earth observation images for downstream analytics.
The sensing functions of multiple satellites are specifically \emph{calibrated} offline, making sure their tiling is aligned, and the overlapping tiles can be uniformly identified across satellites.
This enables multiple satellites to work on the same tile \emph{without actually transmitting its raw data via inter-satellite links}.
As we validate later, this serves to drastically reduce inter-satellite communications by only sharing small-sized intermediate analytics results among collaborating satellites.
A \emph{sensing and analytics pipeline} consists of instances of all analytics functions in a workflow hosted on one or multiple satellites, coupled with the sensing functions on most if not all involved satellites.
Fig.~\ref{fig:orbitchain:sensing_and_analytics} gives an example of such a pipeline for the four analytics functions shown in Fig.~\ref{fig:application} hosted across two satellites.
At time $t_1$, the first satellite's sensing function captures a ground track frame and divides it into image tiles.
The cloud detection function analyzes these tiles, and forwards the identifiers and masks of the non-cloudy tiles to the next function.
The land use classification function analyzes the same input data along with the masks, and sends updated masks of farmland tiles to the second satellite via an inter-satellite link.
These masks are kept until the second satellite passes over the area at time $t_2 = t_1 + \Delta_s$, and captures the overlapping frame with its own sensing function.
The aligned overlapping tiles are then sent to the waterbody monitoring and crop monitoring functions along with the stored tile masks.
The results are then shared with other satellites (\emph{e.g.} for tip-and-cue), or delivered to ground users via low-power low-bandwidth satellite-ground communication channels like LoRa~\cite{gadre2022lowlatency,zhao2025b2lora} or 5G links~\cite{liu2024democratizing}.

To estimate end-to-end workload a pipeline incurs on each analytics function, we utilize the distribution ratios defined in Section~\ref{sec:system:application_abstraction}.
Let $N_0$ be the number of tiles on an original ground track frame.
The average number of tiles that a function $m_{i'}$ needs to process depends on the distribution ratios of all upstream functions of $m_{i'}$ and their function calls, and can be computed by simply traversing the workflow graph from the source.
For instance, in Fig.~\ref{fig:abstraction}, the average number of tiles processed by $m_2$ would be $0.5\cdot N_0$, while the average number processed by both $m_3$ and $m_4$ is $0.25\cdot N_0$.
We use a \emph{workload factor} $\rho_i \ge 0$ to denote the average fraction of tiles that function $m_i$ receives per input tile at the source, where $\rho_1 = 1$, $\rho_2 = 0.5$, $\rho_3 = \rho_4 = 0.25$ in Fig.~\ref{fig:abstraction}.
Note that the factor stays the same even if the number of input tiles changes at the source.
The algorithm for computing $\{ \rho_i \}$ is presented in Appendix~\ref{appdx:real_graph_capacity}.

\textbf{\textit{Remark:}}
For certain satellite orbit parameters, multiple leader-follower satellites may not have exactly overlapping ground tracks, leading to a small fraction of tiles that can only be captured by some but not all satellites in a constellation.
We address this by carefully designing the pipelines and scheduling workloads in Section~\ref{sec:framework}, making sure such data is analyzed by the subset of satellites that can capture it.%

\subsection{Analytics Function Profiling and Performance Modeling}
\label{sec:system:profiling}
\begin{figure*}[th]
\centering
\subfloat[CPU data processing speed.]{\includegraphics[width=0.23\textwidth]{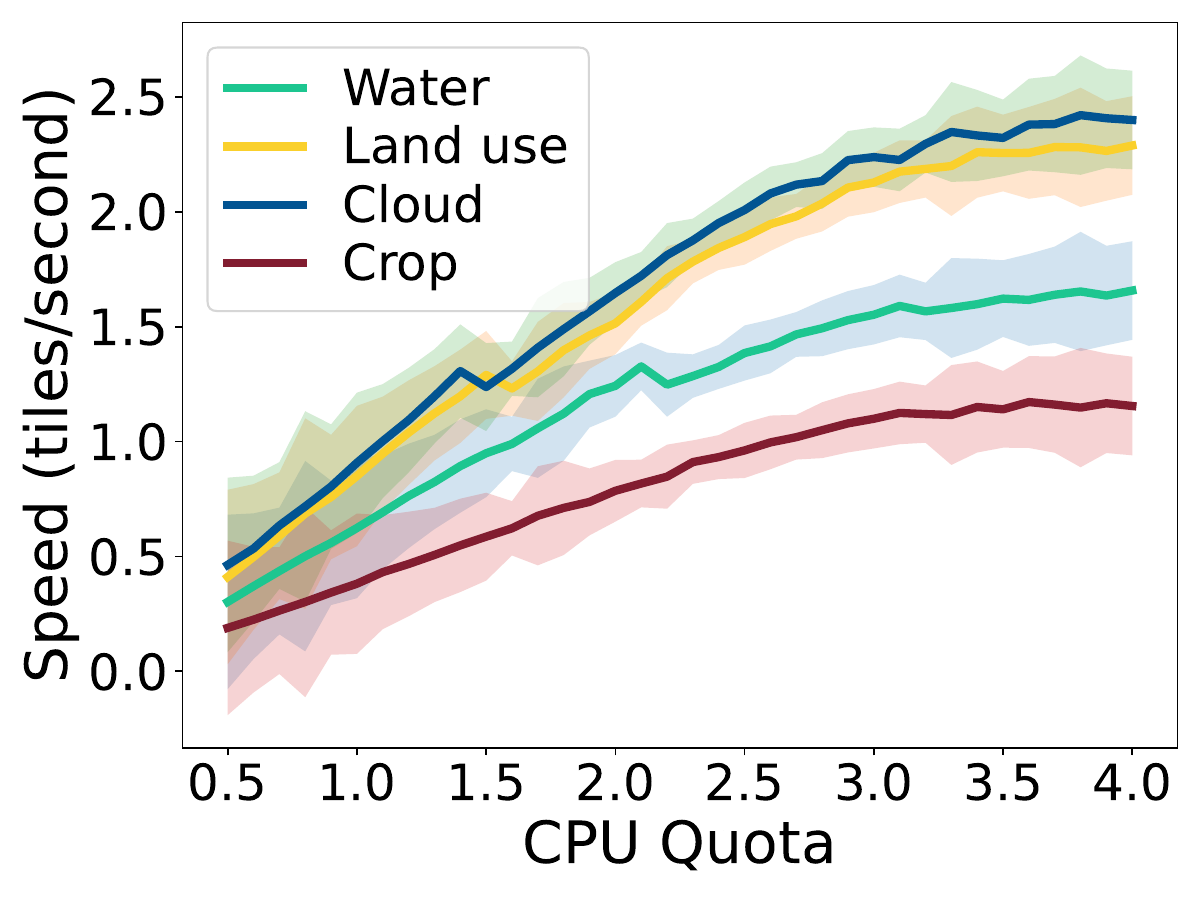}
\label{fig:profiling:cpu_quota_speed_cpu}}
\hfil
\subfloat[GPU data processing speed.]{\includegraphics[width=0.23\textwidth]{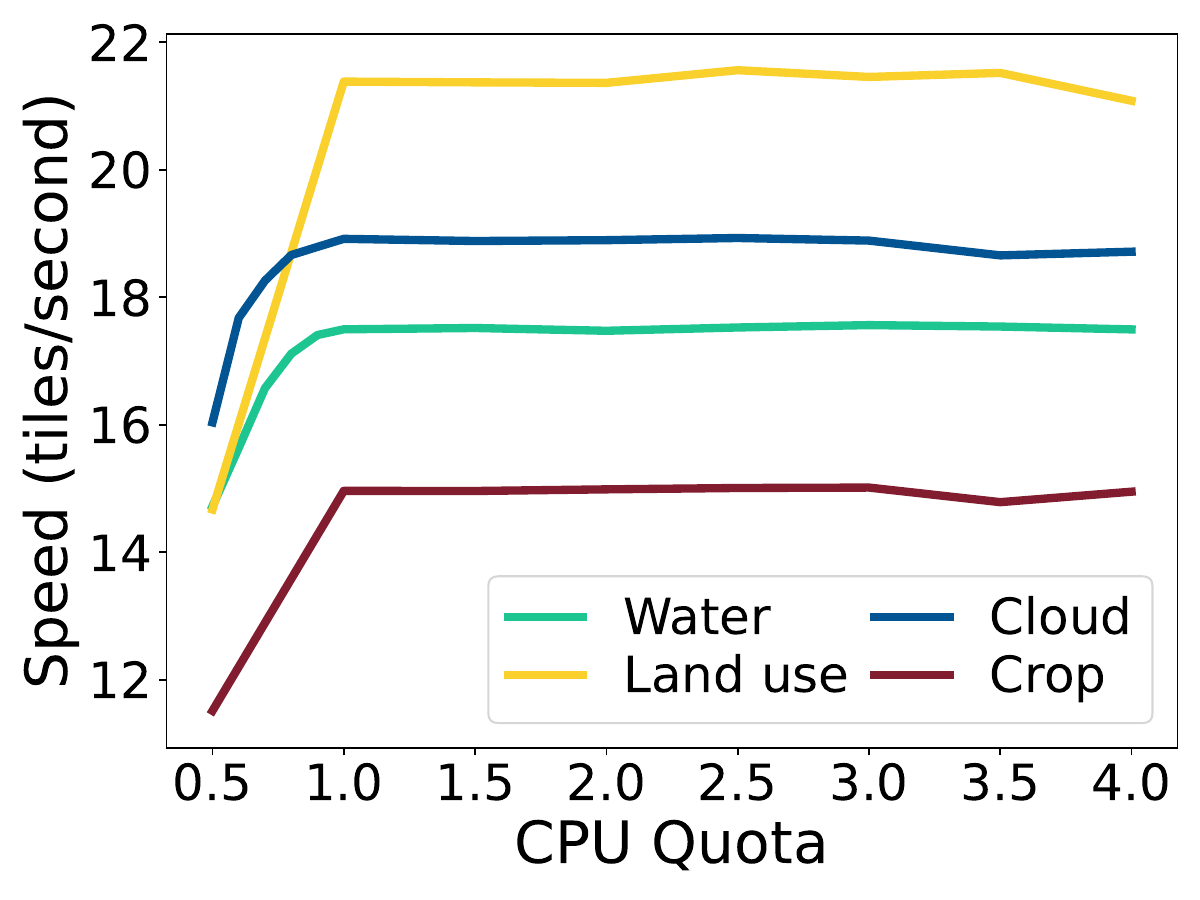}
\label{fig:profiling:cpu_quota_speed_gpu}}
\hfil
\subfloat[CPU memory consumption.]{\includegraphics[width=0.23\textwidth]{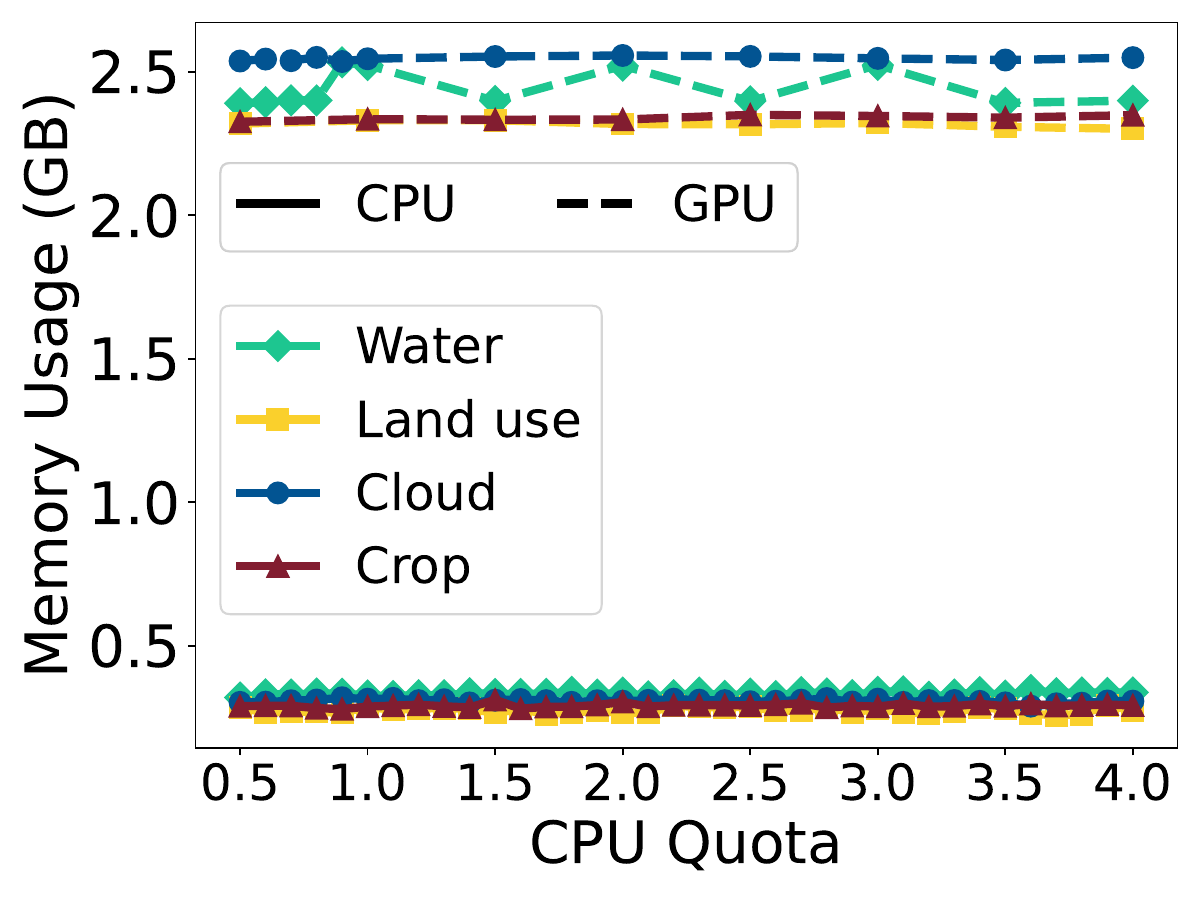}
\label{fig:profiling:cpu_quota_ram_cpu}}
\hfil
\subfloat[Power consumption.]{\includegraphics[width=0.23\textwidth]{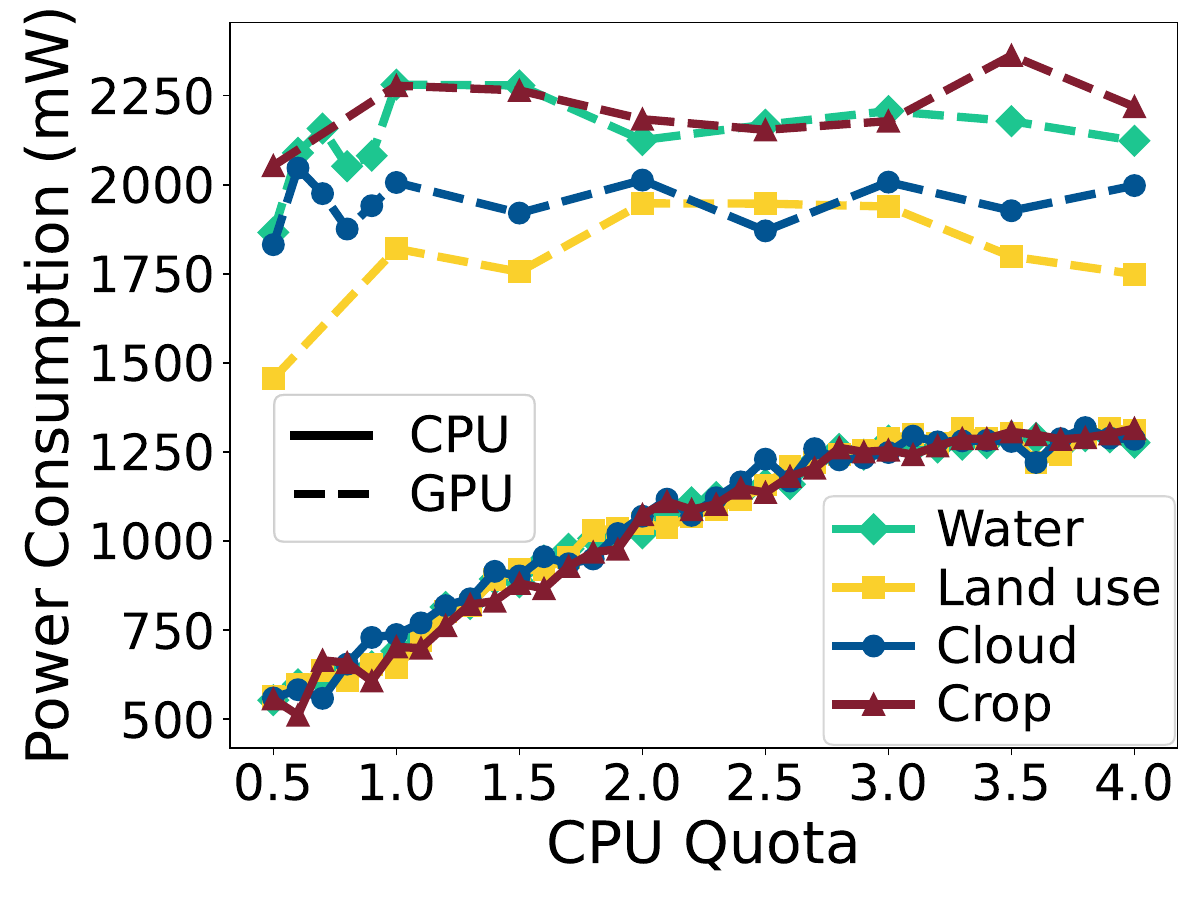}
\label{fig:profiling:power_consumption}}
\hfil
\vspace{-1em}
\caption{Analytics function profiling results. Curves and shadows are the average and standard deviation of profiling results in three rounds. (Cloud: cloud detection; Land use: land use classification; Water: waterbody monitoring; Crop: crop monitoring.)}
\label{fig:profiling}
\end{figure*}

\noindent
Effective pipeline design and resource orchestration relies on accurately modeling the performance and resource utilization of analytics functions on heterogeneous CPU-GPU platforms.
Here we demonstrate profiling and modeling insights of analytics functions for power-constrained devices.%
We profile four deep learning analytics models as in Fig.~\ref{fig:application} across three major analytics tasks~\cite{ultralytics,tan2019efficientnet,sandler2018mobilenetv2}: segmentation, classification, and object detection.
We focus on four critical resources: CPU, GPU, memory, and energy.
We omit storage capacity since existing satellites already have multi-TB storage~\cite{landsat}, sufficient for both temporary data buffering and supporting many analytics functions \emph{as long as the processing speed keeps up with data generation rate (i.e.\ no bufferbloat)}.
Each function undergoes three profiling rounds with varying numbers of CPU cores (CPU quota), with and without GPU acceleration on an orbital edge device (Jetson Orin Nano).
The device is configured to operate at $7$ Watts, corresponding to the solar power input of a standard 3U CubeSat~\cite{3u}.

\textbf{Notations.}
We use set $\mathbf{M}=\{m_1, m_2, \dots, m_{N_m}\}$ to denote the set of $N_m$ analytics functions, with indices topologically sorted according to the workflow graph.
We use set $\mathbf{S}=\{s_1, s_2, \dots, s_{N_s}\}$ to denote the $N_s$ satellites in the constellation, with indices sorted by satellite movement order.

\textbf{CPU and analytics speed.}
We profile how CPU quota affects data analytics speed.
Define $v^{\text{cpu}}_{i, j}$ as the number of tiles per second that can be processed by analytics function $m_i$ using CPU-only execution.
From Fig.~\ref{fig:profiling}\subref{fig:profiling:cpu_quota_speed_cpu}, we observe that $v^{\text{cpu}}_{i, j}$ increases with allocated CPU resources until the device becomes saturated, but the increase is not linear.
Let $r^{\text{cpu}}_{i, j}$ denote the CPU quota allocated to $m_i$ on satellite $s_j$.
We model the CPU-based analytics speed of $m_i$ on satellite $s_j$ as a piecewise linear function $g^{\text{cspeed}}_{i,j}$:
\begin{equation}
    \label{eq:cpu_speed}
    v^{\text{cpu}}_{i, j} = g^{\text{cspeed}}_{i,j} \left(r^{\text{cpu}}_{i, j}\right),
\end{equation}
which balances between accuracy and complexity.
Appendix~\ref{appdx:cpu_speed_fitting} provides more details on the model fitting process based on profiling data.
\textbf{GPU and analytics speed.}
GPUs are available on orbital edge devices like the NVIDIA Jetsons. 
From Fig.~\ref{fig:profiling}\subref{fig:profiling:cpu_quota_speed_gpu}, we observe that GPU-accelerated analytics functions maintain relatively constant data processing speeds once sufficient CPU quota is allocated.
GPU acceleration is necessary and can achieve $10$-$20\times$ faster speed than CPU-only processing even under constrained power.
We model the GPU inference speed as a constant $v^{\text{gpu}}_{i,j}$ for analytics function $m_i$ on satellite $s_j$, when the basic CPU quota for full-speed GPU analysis, denoted as $r^{\text{gcpu}}_{i}$, has been allocated.

\textbf{Memory utilization.}
As discussed in Section~\ref{sec:preliminary_study:oec}, analytics functions are terminated when on-board memory is exhausted.
We profile peak memory usage to determine a function's minimum memory requirement.
Results are shown in Fig.~\ref{fig:profiling}\subref{fig:profiling:cpu_quota_ram_cpu}.
Peak memory usage remains stable across devices and does not vary with the allocated CPU quota.
We denote this peak usage for analytics function $m_i \in \mathbf{M}$ as $r^{\text{cmem}}_{i}$ for CPU-based inference and $r^{\text{gmem}}_{i}$ for GPU-based inference.

\textbf{Power consumption.}
We profile CPU and GPU inference power consumption, as shown in Fig.~\ref{fig:profiling}\subref{fig:profiling:power_consumption}.
Across all analytics functions, GPU-accelerated inference consumes over $1.5\times$ the power of CPU inference.
CPU inference power increases monotonically with CPU quota, while GPU inference power remains nearly constant once sufficient CPU quota is allocated.
We model CPU inference power consumption for analytics function $m_i$ on satellite $s_j$ as
\begin{equation}
    \label{eq:cpu_power}
    r^{\text{cpow}}_{i, j} = g^{\text{cpow}}_{i,j} \left(r^{\text{cpu}}_{i, j}\right).
\end{equation}
where $g^{\text{cpow}}_{i,j}$ is a piecewise linear function similar to $g^{\text{cspeed}}_{i,j}$.
If available, the GPU inference power consumption, after allocating sufficient CPU quota, is denoted as $r^{gpow}_{i,j}$. %

\begin{figure}[t]
\vspace{-12pt}
\centering
\subfloat[GPU inference cold start.]{\includegraphics[width=0.225\textwidth]{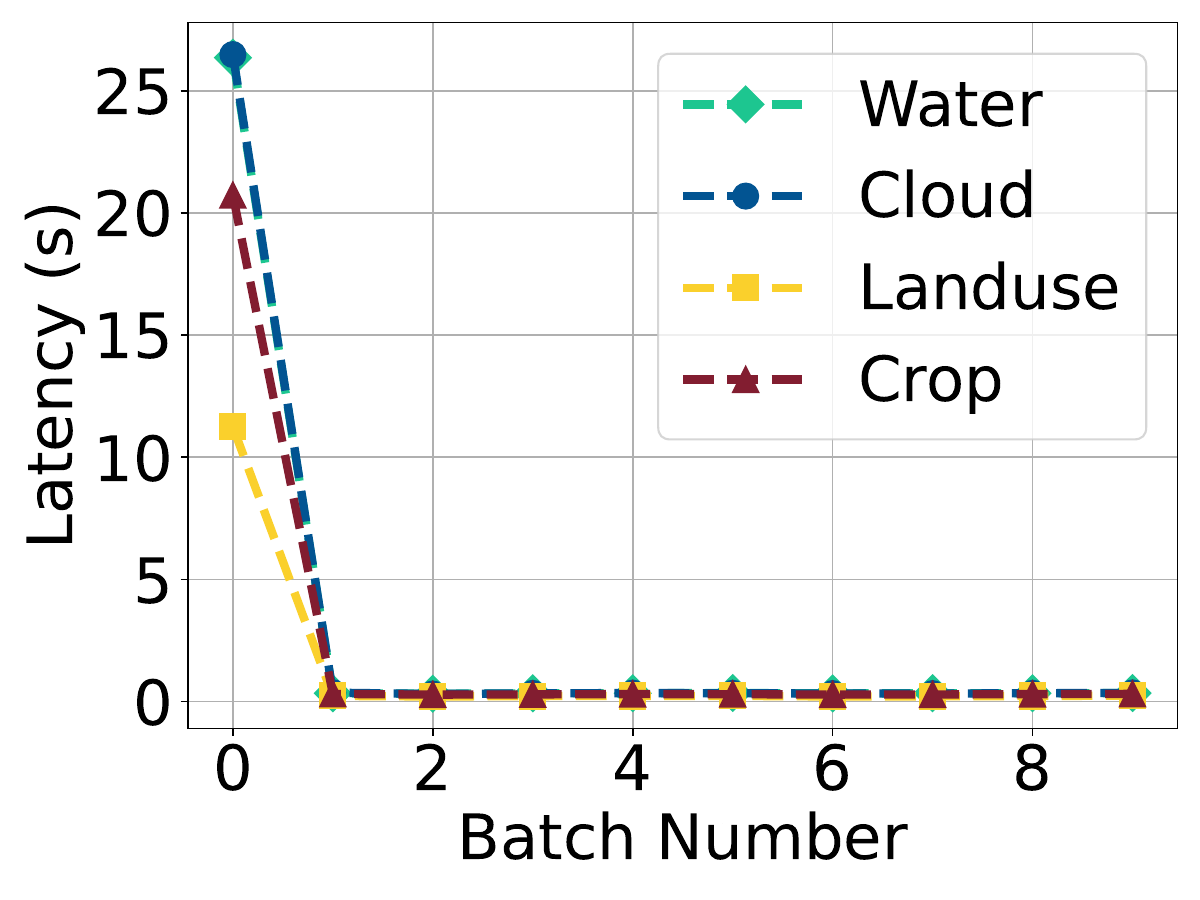}
\label{fig:profiling_props:cold_start}}
\hfil
\subfloat[Per-frame analytics data size.]{\includegraphics[width=0.225\textwidth]{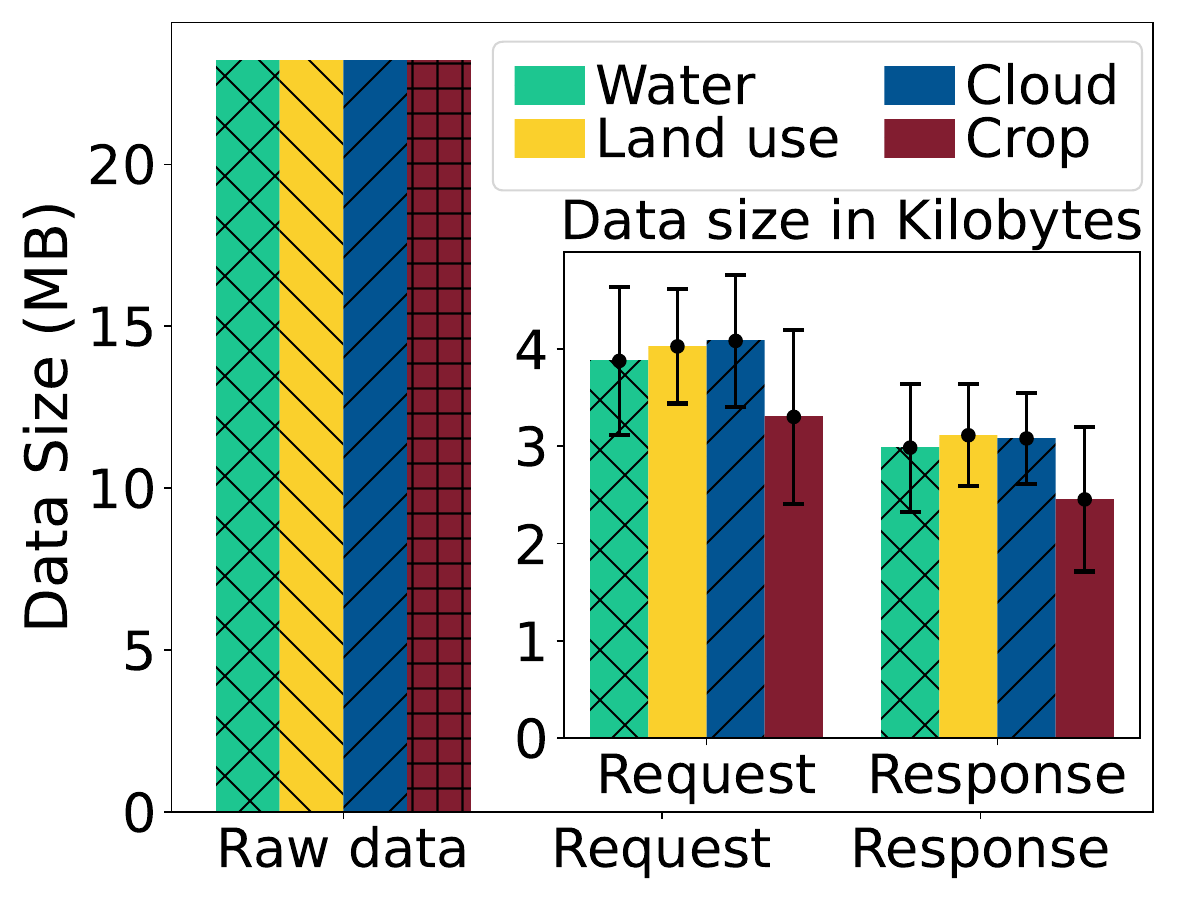}
\label{fig:profiling_props:communication}}
\hfil
\vspace{-1em}
\caption{(a) GPU inference latency with rounds of inference. (b) Data size of sending different data types.}
\label{fig:profiling_props}
\end{figure}

\textbf{Cold start overhead for GPU inference.}
GPU inference typically incurs cold-start overhead due to model loading.
If a model is instantiated only when an incoming frame becomes available, the first batch of tiles will incur a significant delay compared to the rest of the frame, as shown in Fig~\ref{fig:profiling_props}\subref{fig:profiling_props:cold_start}.
\textbf{Inter-satellite communication overhead.}
We profile the data volume flowing in and out of each analytics function and present the results in Fig.~\ref{fig:profiling_props}\subref{fig:profiling_props:communication}.
The key insight is that intermediate analytics results (\emph{e.g.}\ detected objects, segmentation masks, etc.) are 5-6 orders of magnitude smaller than raw sensing data sizes. %
This validates the potential savings in communication power consumption and bandwidth for utilizing local sensing functions in pipelined processing instead of exchanging raw data via inter-satellite links.
\subsection{Design Insights from Profiling}
\label{sec:principles}

\noindent
The profiling results give or validate several key insights that drive the design of {\oursol} below:
\begin{enumerate}[leftmargin=1.5em]
    \item \textbf{Data locality:} Due to huge differences between raw data volumes and intermediate results, pipelines should be designed to maximally utilize local sensing functions and avoid inter-satellite raw data communications.
    \item \textbf{CPU-GPU collaboration:} While edge GPU has a huge processing speed advantage over CPU, the former cannot be easily parallelized among analytics functions and also consumes more energy. CPU-GPU collaborative processing is necessary to balance the processing throughput of different functions and control the power consumption.
    \item \textbf{Model instantiation:} To avoid cold-start in inference, models should be loaded and continually operating under power and resource constraints to minimize processing latency and avoid buffer buildup.
\end{enumerate}

\section{{\oursol} Design}
\label{sec:framework}

\noindent
{\oursol} is a real-time Earth observation framework that can deliver analytics results for time-sensitive tasks in minutes, by enabling efficient multi-satellite collaborative in-orbit analytics of Earth sensing data.
In {\oursol}, multiple leader-follower satellites jointly instantiate sensing and analytics pipelines, and balance the overall analytics workload via efficient analytics function deployment, resource allocation and workload routing.
As its main objective, {\oursol} ensures highly efficient frame processing within the frame deadline, thus achieving long-term stability of the on-board buffer and preventing data loss caused by buffer overload.

Below, we start with an overview of {\oursol}, followed by detailed design of each component.

\subsection{{\oursol} Overview}

\begin{figure}[!t]
\centering
\includegraphics[width=0.47\textwidth]{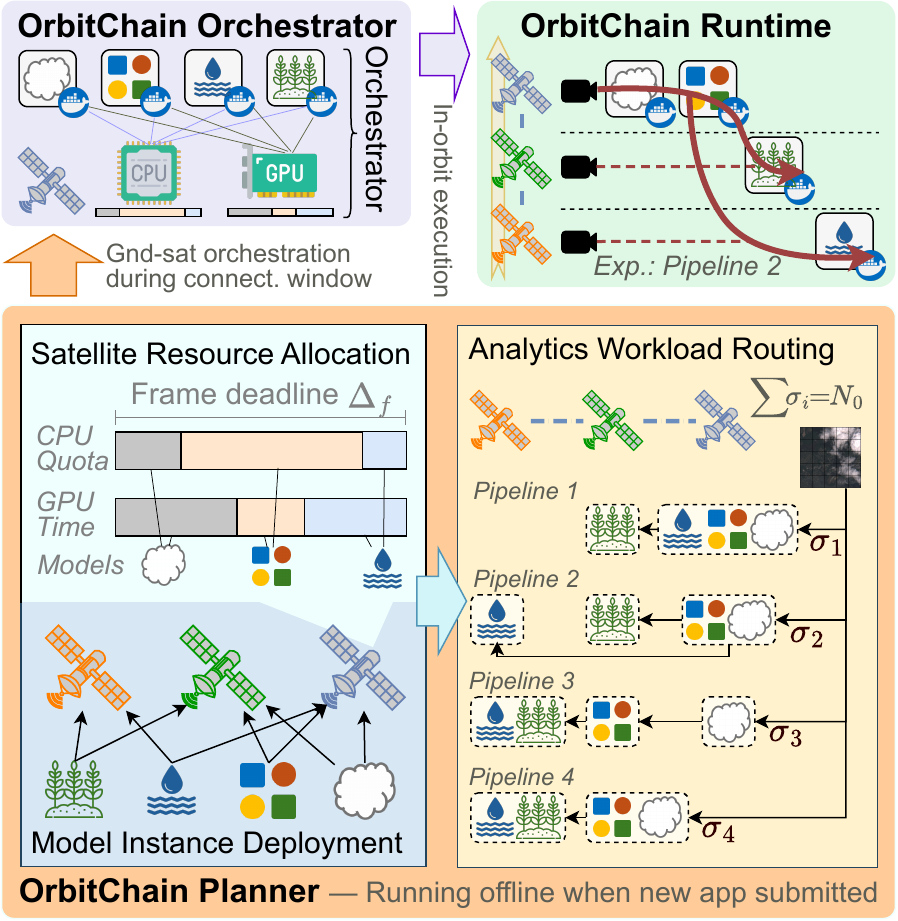}
\vspace{-1em}
\caption{ {\oursol} workflow. The Planner processes analytics requests on the ground, and decides in-orbit deployment, resource allocation and workload routing. The Orchestrator instantiates containerized functions via ground-satellite control. The Runtime module on each satellite dynamically routes input data through instantiated pipelines to generate real-time insights.}
\label{fig:workflow}
\end{figure}

\noindent
The overall workflow of {\oursol} is shown in Fig.~\ref{fig:workflow}, which has three phases: planning, orchestration, and runtime.

\textbf{Planning.}
Planning is performed on the ground when there is a change in the analytics workflow or the constellation.
For a new user analytics workflow request, the ground command center computes the best option for instantiating the workflow with sensing and analytics pipelines, and decides resource allocation and workload routing based on offline profiling results.
This phase is performed offline based on predicted satellite orbits, resources and energy inputs, and the currently scheduled workloads in the constellation.
\textbf{Deployment.}
Once the planning phase is completed, the ground command center communicates with the constellation to instantiate the assigned sensing and analytics pipelines.
A ground-satellite orchestrator performs this job, instantiating containerized analytics functions and orchestrating workload steering among them.
Containerization allows flexible instantiation as well as  resource control of the functions~\cite{denby2023kodan}.
For instance, Docker allows CPU quota allocation for containers via their \texttt{cpu\_quota} and \texttt{cpu\_period} attributes~\cite{docker_resource_constraints}.
Meanwhile, GPU time slicing is achieved at runtime based on pre-defined model switching time schedules during orchestration.
The orchestration can usually be performed in real-time using the standard Telemetry, Tracking and Command (TT\&C) communication channels.
However, if a model is user-supplied and not already available on the satellites, it needs to be uploaded to the satellites at the next high-speed ground-satellite connection.
See Appendix~\ref{appdx:implementation:constellation_control} for more details on constellation control.

\textbf{Runtime.}
At runtime, analytics functions across satellites collaborate to execute sensing and analytics pipelines based on incoming data.
At each sensing function, each input tile is tagged with a specific sensing and analytics pipeline, and routed towards downstream function instances based on the tag.
When a cross-satellite function call is made from another satellite, tags of the remote call data (intermediate results) and the local tiles will be matched, to ensure each tile is processed with the correct inputs.
Meanwhile, an online scheduler manages GPU time slicing on each satellite, rotating through analytics functions based on the pre-defined time table.
Each analytics function thus gets a pre-allocated share of the GPU processing time to speed-up its local processing of its input data queue.
At the end of a sensing and analytics pipeline, final analytics results are delivered to another satellite for tip-and-cue, or to the end user again via TT\&C or another constrained channel like low-power LoRa.%
Below, we focus on the ground planning phase which is the core engine behind efficient constellation operations.
The other two phases can be implemented via fairly standard containerization and orchestration tools, and thus will be omitted except in evaluation.

\subsection{Analytics Function Deployment and Resource Allocation}
\label{sec:framework:deployment}

\noindent
{\oursol} adopts a pipelined model, where each analytics function continuously processes its input data stream from upstream functions and forwards its results to downstream ones.
Each function (except sensing) thus maintains an input queue of input data, and continuously processes the queue until it is empty.
Because a frame is generated per frame deadline at the source, each function must finish analyzing all tiles of an incoming frame before the next one arrives, in order to ensure overall queue stability.
Meanwhile, each function requires a certain amount of resources and energy budget to operate based on the profiling results.
We formulate this as a joint analytics function deployment and resource allocation problem, constrained by resource and energy budgets as well as the per-frame processing deadline.%
\textbf{Decision variables.}
For function deployment, we define $x_{i, j} \in \{ 0, 1\}$ as deploying an instance of function $m_i$ on satellite $s_j$.
The CPU quota allocation to $m_i$ on satellite $s_j$ is then denoted by $r^{\text{cpu}}_{i, j} \ge 0$.
Meanwhile, consider GPU acceleration of certain functions on certain satellites, we define $y_{i, j} \in \{ 0, 1\}$ as whether $m_i$ could enjoy GPU acceleration on satellite $s_j$.
The GPU is time-sliced within each frame deadline, and we let $t_{i, j} \ge 0$ denote the time slice allocated to $m_i$ within one frame deadline at satellite $s_j$.
We denote the four sets of variables as 
$\mathbf{X} = \{ x_{i, j} \}_{i,j}$,
$\mathbf{R} = \{ r^{\text{cpu}}_{i, j} \}_{i,j}$,
$\mathbf{Y} = \{ y_{i, j} \}_{i,j}$, and
$\mathbf{T} = \{ t_{i, j} \}_{i,j}$, respectively.

\textbf{Total workload constraint.}
To prevent workload accumulation in queues, the core constraint is that for every analytics function, the number of tiles that can be analyzed within a frame deadline by all instances must be no less than the total number of tiles in a frame $N_0$ times the workload factor ratio $\rho_i$ defined in Section~\ref{sec:sense_ana_pipe}:
\begin{equation}
    \label{eq:constraint:speed}
    \sum_{s_j \in \mathbf{S}} \left( v^{\text{cpu}}_{i, j} \cdot \Delta_f + v^{\text{gpu}}_{i, j} \cdot t_{i,j} \right) \ge \rho_i \cdot N_0, \;\;\forall m_i \in \mathbf{M}.
\end{equation}
Recall that $v^{\text{cpu}}_{i, j} = g^{\text{cspeed}}_{i,j} (r^{\text{cpu}}_{i, j})$ while $v^{\text{gpu}}_{i, j}$ is a constant based on profiling.

\textbf{CPU and GPU allocation.}
For CPU, the allocation for a function consists of the CPU quota allocated for CPU-only execution, plus the minimum CPU quota allocated for GPU acceleration if GPU is assigned.
The total CPU quota on satellite $s_j$ is bounded by the number of CPU cores $c^{\text{cpu}}_{j}$, times a safety margin $\beta \in (0, 1)$ where the remaining $(1-\beta)$ is left for safety-critical background processes:
\begin{equation}
    \label{eq:constraint:cpu}
    \sum_{m_i \in \mathbf{M}} \left( r^{\text{cpu}}_{i,j} + r^{\text{gcpu}}_{i} \cdot y_{i, j} \right) \le \beta \cdot c^{\text{cpu}}_{j}, \;\;\forall s_j \in \mathbf{S}.
\end{equation}
For GPU sharing, the total GPU time slices must be bounded by a fraction of the frame deadline $\Delta_f$:
\begin{equation}
    \label{eq:constraint:time_slice}
    \sum_{m_i \in \mathbf{M}} t_{i,j} \le \alpha\Delta_f, \;\;\forall s_j \in \mathbf{S},
\end{equation}
where $\alpha \in (0, 1)$ is a context-switching discount coefficient.

\textbf{Minimum CPU quota and GPU slice.}
From profiling, we notice that each analytics function requires a minimum CPU quota $lb^\text{cpu}_{i}$ to be instantiated. 
For our example analytics functions, the minimum CPU quotas are $0.5$, which is why Fig.~\ref{fig:profiling} start from CPU quota of $0.5$.

\begin{equation}
    \label{eq:constraint:min_quota}
    r^\text{cpu}_{i,j} \ge lb^\text{cpu}_{i} \cdot x_{i,j}, \;\;\forall m_i \in \mathbf{M}, s_j \in \mathbf{S}.
\end{equation}
While GPU time slicing does not have a similar limit, we notice that frequent context switching between functions cause non-negligible overhead that slows down processing.
Hence to avoid frequent context switching, we require each function's GPU time slice to have a minimum length $lb^\text{gpu}_{i}$:
\begin{equation}
    \label{eq:constraint:min_gpu_quota}
    t_{i,j} \ge lb^\text{gpu}_{i} \cdot y_{i,j}, \;\;\forall m_i \in \mathbf{M}, s_j \in \mathbf{S}.
\end{equation}

\textbf{Memory allocation.}
Instantiated analytics functions consume a nearly constant amount of memory when operating.
In a shared-memory architecture (like NVIDIA Jetson) where CPU and GPU share the same memory, the total memory consumption of both CPU and GPU instances is bounded:
\begin{equation}
    \label{eq:constraint:memory}
    \sum_{m_i \in \mathbf{M}} \left( r^\text{cmem}_{i} \cdot x_{i,j}  + r^\text{gmem}_{i} \cdot y_{i, j} \right) \le c^\text{mem}_{j}, \;\;\forall s_j \in \mathbf{S}.
\end{equation}
The constraint can be separated into two if CPU and GPU each has its own memory.

\textbf{Power constraint.}
Satellite operates on solar power.
The total power consumption on satellite $s_j$ cannot exceed the available power budget $c^\text{pow}_{j}$ for on-board analytics:
\begin{equation}
    \label{eq:constraint:energy}
    \sum_{m_i \in \mathbf{M}} r^{\text{cpow}}_{i, j} + \max_{m_i \in \mathbf{M}} \left( r^{\text{gpow}}_{i, j} \cdot y_{i,j} \right) \le c^{\text{pow}}_{j}, \;\;\forall s_j \in \mathbf{S}.
\end{equation}
We use the maximum GPU-acceleration power consumption across all functions, since the GPU is time-sliced and only one analytics function is actively running on it at a time.

\textbf{Formulation and complexity.}
The problem of analytics function deployment and resource allocation is formulated as the following Mixed-Integer Linear Program (MILP):
\begin{equation}
    \label{eq:objective}
    \begin{aligned}
        & \text{find } \{\mathbf{X}, \mathbf{R}, \mathbf{Y}, \mathbf{T}\}, \;
        & \text{subject to } \eqref{eq:constraint:cpu}\text{--}\eqref{eq:constraint:energy}.
    \end{aligned}
\end{equation}
It ensures that jointly, all function instances have sufficient capacity to complete the processing of all tiles within a frame deadline, thus guaranteeing long-term stability of the on-board buffer.
While Eq.~\eqref{eq:objective} does not specify an \emph{objective}, the formulation easily admits various operator goals, such as minimizing energy consumption or load balancing across satellites.
Our implementation maximizes the bottleneck capacity among all function instances to reduce the impact of temporary performance fluctuation.

The MILP has $2 \times N_m \times N_s$ variables.
Because this problem is solved only once per workflow change during offline planning, the problem can be relatively efficiently solved by a commercial solver such as Gurobi~\cite{gurobi}.
Some discussions about planning efficiency is further attached in Appendix~\ref{appdx:implementation:planning}.

\begin{figure}[t]
\centering
\includegraphics[width=0.35\textwidth]{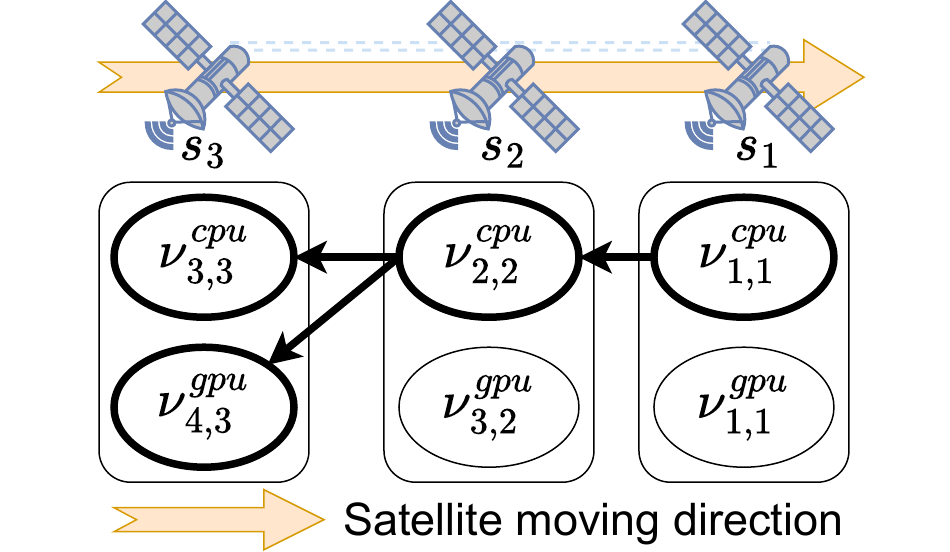}
\vspace{-1em}
\caption{A pipeline generated by workload routing.}
\label{fig:real_graph}
\end{figure}

\subsection{Analytics Workload Routing}
\label{sec:framework:routing}
Deployed analytics functions must collaborate to execute Earth observation analytics workflows.
To achieve this, the deployed function instances must be orchestrated into sensing and analytics pipelines.  %
The process of analytics workload routing involves deciding the unique end-to-end pipeline that each tile needs to be processed with.
In this process, a major consideration is how to minimize the inter-satellite communication overhead incurred by cross-satellite function calls.
As we illustrate below, this requires proper modeling of the end-to-end processing capacity of a target pipeline, and intelligent workload division among different pipelines.
\textbf{Function instance capacity.}
Consider an instance of function $m_i \in \mathbf{M}$ deployed on a satellite $s_j \in \mathbf{S}$.
Its processing speed $v^{\text{cpu}}_{i,j}$ is decided by the allocated CPU quota $r^{\text{cpu}}_{i,j}$.
If GPU acceleration is available and assigned, $m_i$ could instead gain $v^{\text{gpu}}_{i,j}$ speed during the accelerated time.
To simplify modeling, we regard CPU-only and GPU-accelerated execution of $m_i$ as two different instances, denoted as $\nu^\text{cpu}_{i,j}$ and $\nu^\text{gpu}_{i,j}$ respectively.
Define the \emph{capacity} of instance $\nu^d_{i,j}$ as:
\begin{equation}
    \label{eq:instance_capacity}
    n^d_{i,j} =
        \begin{cases}
            v^\text{cpu}_{i, j} \cdot \Delta_f, & \text{if } d = {\text{cpu}}, \\
            v^\text{gpu}_{i, j} \cdot t_{i,j}, & \text{if } d = {\text{gpu}},
        \end{cases}
\end{equation}
where $d \in \{ \text{cpu}, \text{gpu} \}$ is a device indicator.
The capacity is in the unit of \emph{number of incoming tiles per frame deadline}.

\textbf{Pipeline capacity.}
Consider now a sensing and analytics pipeline $\zeta_k$ that consists of a set of function instances across multiple satellites.
For rigor of discussion, we restrict each pipeline to have exactly \emph{one} instance $\nu^d_{i,j}$ per function $m_i$.
The capacity of the pipeline is generally decided by the bottleneck capacity across all function instances.
However, because each function receives a different number of incoming tiles for a source frame, the workload incurred on each function per frame deadline is different.
Hence we use the workload factor $\rho_i$ computed in Section~\ref{sec:sense_ana_pipe} to define the pipeline capacity $\sigma_k$, in the unit of \emph{number of source tiles per frame deadline}:
\begin{equation}
    \label{sec:pipe_cap}
    \sigma_k = \min_i \frac{n^{d_{i}^k}_{i,j}}{\rho_i}.
\end{equation}
Note that $d_i^k \in \{ \text{cpu}, \text{gpu} \}$ is the device used by the instance of $m_i$ in this pipeline $\zeta_k$.

\textbf{Analytics workload routing algorithm.}
Algorithm~\ref{alg:routing} shows our algorithm for routing analytics workloads across deployed function instances.
The outer loop in Lines~\ref{alg:outer_start}--\ref{alg:outer_end} iterates to find new pipelines with available capacity until all tiles in the source frame is processed.
Within the outer loop, the algorithm employs a Breath-First Search (BFS) to search for a pipeline with available capacity (Line~\ref{alg:routing:assign_head}).
It starts from a dummy instance $\nu^0_{0, 0}$, which connects to an instance of each function with in-degree of $0$ in the workflow graph, on the \emph{first} satellite where such an instance has a \emph{positive} remaining capacity.
For each downstream function $m_{i'}$, the algorithm looks for the instance $n^{d^*}_{i',j^*}$ that has available capacity and is within the minimum number of hops from the current instance's satellite (Lines~\ref{alg:routing:start_find_min_hop}--\ref{alg:routing:end_find_min_hop}), and add to the current pipeline (Line~\ref{alg:routing:endAddInstance}).
After all functions are assigned an instance in the current $\zeta_k$, the algorithm calculates the pipeline capacity $\sigma_k$ (Line~\ref{alg:routing:getRealGraphCap}), and assign the same amount of workload to $\zeta_k$.
Finally, the algorithm updates the remaining capacities of instances, adjusts connections from the dummy node $\nu^0_{0,0}$ based on updated capacities, and deducts the assigned workload $\sigma_k$ from $N_0$ (Lines~\ref{alg:routing:start_update}--\ref{alg:routing:end_update}).
Overall, the algorithm tries to minimize inter-satellite communication, by each time selecting the downstream instance with the minimum number of hops from the upstream instance.

Let $N_\nu$ be the total number of instances across all satellites, and $N_{\nu,\text{max}}$ as the maximum number of instances per function.
The outer loop runs for at most $N_\nu$ iterations since it saturates at least one function instance per iteration.
Each iteration performs a BFS on a graph that has $N_\nu$ nodes for all the instances and at most $N_{\nu,\text{max}}^2$ links for their upstream-downstream relations.
Overall, Algorithm~\ref{alg:routing} has a time complexity of $O(N_\nu \cdot (N_\nu+ N_{\nu,\text{max}}^2))$.

\begin{algorithm}[!t]
\small
\SetAlgoNoEnd
\caption{\mbox{Analytics workload routing}}
\label{alg:routing}

\KwIn{%
Satellite nodes $\mathbf{S} = \{s_j\}$, 
analytics function instances $\{ \nu^d_{i,j} \}$, 
capacities $\{n^d_{i,j}\}$,
workload factors $\{\rho_i\}$,
source tile number $N_0$.
}
\KwOut{%
Orchestrated pipelines $\mathbf{Z} = \{\zeta_k\,|\,k=1,2,\dots\}$, 
workload allocation $\mathbf{\Sigma} = \{\sigma_k\,|\,k=1,2,\dots\}$.
}  

    $k \leftarrow 1$\;

    \While{$N_0 > 0$}  
    {    \label{alg:routing:begin_add_real_graph}\label{alg:outer_start}
        $\zeta_k \leftarrow \emptyset$, 
        $Q \leftarrow [\nu^0_{0, 0}]$\; \label{alg:routing:assign_head}

        \tcp{BFS to find next available pipeline}
        \While{$Q \ne \emptyset$}{
            $\nu^d_{i,j} \leftarrow Q.pop()$\;%

            \For{$m_{i'} \in downstream(m_i)$}{

                \eIf{$\exists j', d'$ such that $\nu^{d'}_{i', j'} \in \zeta_k$}{ \label{alg:routing:start_find_min_hop}
                    $j^*, d^* \leftarrow j', d'$\;
                }{
                
                    $j^*, d^* \leftarrow \arg\min_{j', d'} \{ |j' - j| \text{ s.t. } n^{d'}_{i',j'} \!>\! 0 \}$\; \label{alg:routing:end_find_min_hop}
                }

                \If{$j^*, d^* = \emptyset, \emptyset$}{
                    \Return{Infeasible.}
                }

                $\zeta_k \leftarrow \zeta_k \cup \{\nu^{d^*}_{i', j^*}\}$\; \label{alg:routing:endAddInstance}
    
                $Q.push( \nu^{d^*}_{i', j^*} )$\;

            }

            \label{alg:routing:end_add_to_real_graph}
        }

        $\sigma_k \leftarrow \min \left\{ n^{d}_{i, j} / \rho_i \,|\, \forall \nu^d_{i,j} \in \zeta_k \right\}$\; \label{alg:routing:getRealGraphCap}

        $\mathbf{Z} \leftarrow \mathbf{Z} \cup \zeta_k$, $\mathbf{\Sigma} \leftarrow \mathbf{\Sigma} \cup \sigma_k$\;

        \For{$\nu^d_{i,j} \in \mathbf{V}_k$} {  \label{alg:routing:start_update}
            $n_{i,j}^d \leftarrow n_{i,j}^d - \sigma_k \cdot \rho_i$\; %
        }

        Update the downstream instance(s) of $\nu^0_{0,0}$\;

        $N_0 \leftarrow N_0 - \sigma_k$, $k \leftarrow k + 1$\;
        \label{alg:routing:end_update}
        
    } \label{alg:outer_end}
    
    \Return{$\mathbf{Z}$, $\mathbf{\Sigma}$.}
\end{algorithm}

\subsection{Handling Ground Track Shifts}

\noindent
As mentioned in Section~\ref{sec:sense_ana_pipe}, ground tracks of leader-follower satellites may not exactly align with each other due to natural orbit formation.
When it happens, some tiles may only be captured by a subset of satellites, and other satellite would not be able to process this tile using the local sensing function.
We address this by ensuring that these tiles can processed by the set of satellites that can capture them.

\textbf{Deployment and resource allocation:}
Let $\mathbf{\bar S} \subseteq \mathbf{S}$ be a set of satellites capturing a subset of tiles $\mathbf{I}_{\mathbf{\bar S}}$ that cannot be captured by other satellites in $\mathbf{S} \setminus \mathbf{\bar S}$.
Because satellite orbit shifts are contiguous in a sequence of leader-follower satellites, there are at most ${|\mathbf{S}|\cdot(|\mathbf{S}|+1)}/{2}$ such subsets: $\{ s_1 \}, \{ s_1, s_2 \}, \{ s_1, s_2, s_3 \}, \dots, \{ s_2, s_3 \}, \{ s_2, s_3, s_4 \}, \dots, \{ s_{|\mathbf{S}|} \}$.
One may also consider only the first $|\mathbf{S}|$ of these subsets, corresponding to only the tiles that the leader satellite captures.
In any case, let $\mathcal{S} = \{ \mathbf{\bar S} \}$ be all such subsets under consideration.
In Program~\eqref{eq:objective}, we replace the workload constraint~\eqref{eq:constraint:speed} by a set of constraints, one for each $\mathbf{\bar S} \in \mathcal{S}$:
\begin{equation}
    \label{eq:constraint:speed-new}
    \sum_{s_j \in \mathbf{\bar S}} \left( v^{\text{cpu}}_{i, j} \cdot \Delta_f + v^{\text{gpu}}_{i, j} \cdot t_{i,j} \right) \ge \rho_i \cdot |\mathbf{I}_{\mathbf{\bar S}}|, \;\;\forall m_i \in \mathbf{M}, \mathbf{\bar S} \in \mathcal{S}.
\end{equation}
This constraint ensures that, jointly all satellites in $\mathbf{\bar S}$ have sufficient instances and resources to process all tiles that are uniquely captured by $\mathbf{\bar S}$.
This increases the number of constraints in~\eqref{eq:objective} by at most a factor of $|\mathcal{S}|$ but does not increase the size (number of variables) of the program, and hence has minimal impact on its time complexity.

\textbf{Workload routing:}
In Algorithm~\ref{alg:routing}, we only need to: 
(i) replace the outer loop in Lines~\ref{alg:outer_start}--\ref{alg:outer_end} with $|\mathcal{S}|$ such loops, one for each $\mathbf{\bar S} \in \mathcal{S}$ in increasing order of the size of each $\mathbf{\bar S}$;
(ii) only search for $s_{j'} \in \mathbf{\bar S}$ in Line~\ref{alg:routing:end_find_min_hop}.
This ensures that tiles unique to fewer satellites are assigned a pipeline prior to tiles captured by more satellites, and hence every tile can be processed by satellites capturing it.
This adds a factor of $|\mathcal{S}|$ to the time complexity of Algorithm~\ref{alg:routing}, which is typically small given the limited number of satellites in a constellation.%

\section{Evaluation}
\label{sec:evaluation}
\noindent
In this section, we evaluate our solution on a hardware-in-the-loop OEC testbed with three NVIDIA Jetson Orin Nano devices (CPU+GPU) and four Raspberry Pi Model 4B devices (CPU-only).
Details of the testbed are shown in Appendix~\ref{appdx:testbed}. 
We demonstrate {\oursol}'s ability to complete Earth observation analytics tasks fully in orbit, reduce inter-satellite communication overhead, and achieve real-time analytics.%

\begin{figure*}[th]
\centering
\subfloat[Cloud--land use--crop.]{\includegraphics[width=0.23\textwidth]{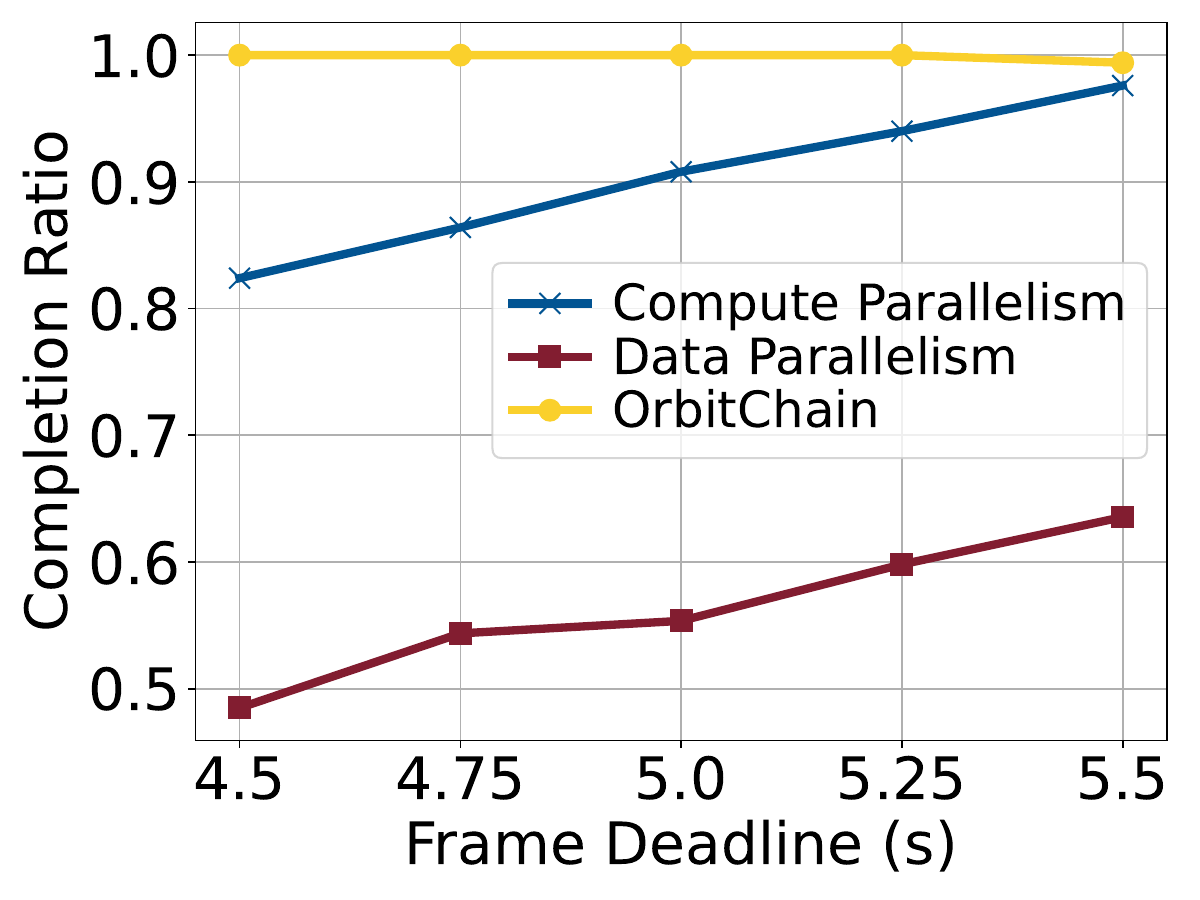}
\label{fig:completion_ratio:jetson_cloud_landuse_object}}
\hfil
\subfloat[Cloud--land use--water.]{\includegraphics[width=0.23\textwidth]{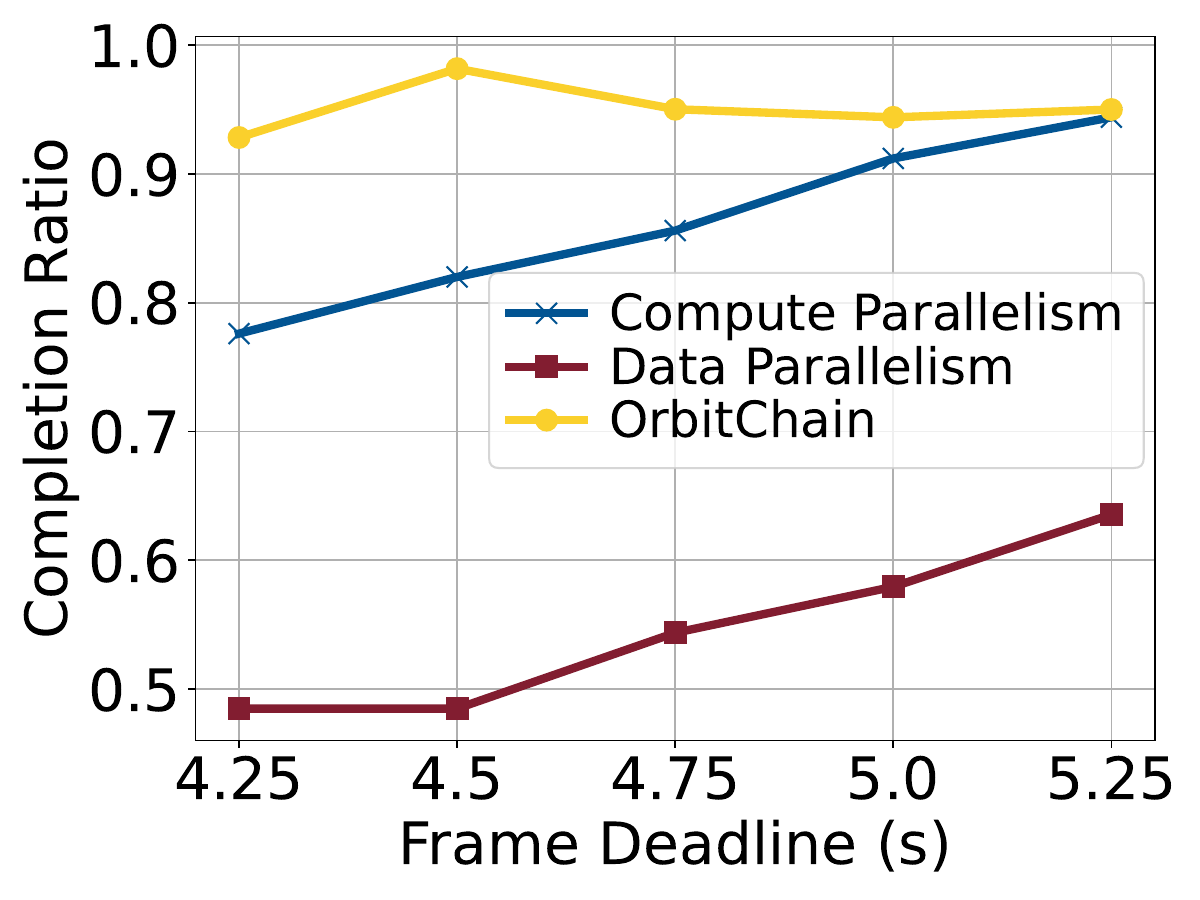}
\label{fig:completion_ratio:jetson_cloud_landuse_water}}
\hfil
\subfloat[Cloud{--}crop and water.]{\includegraphics[width=0.23\textwidth]{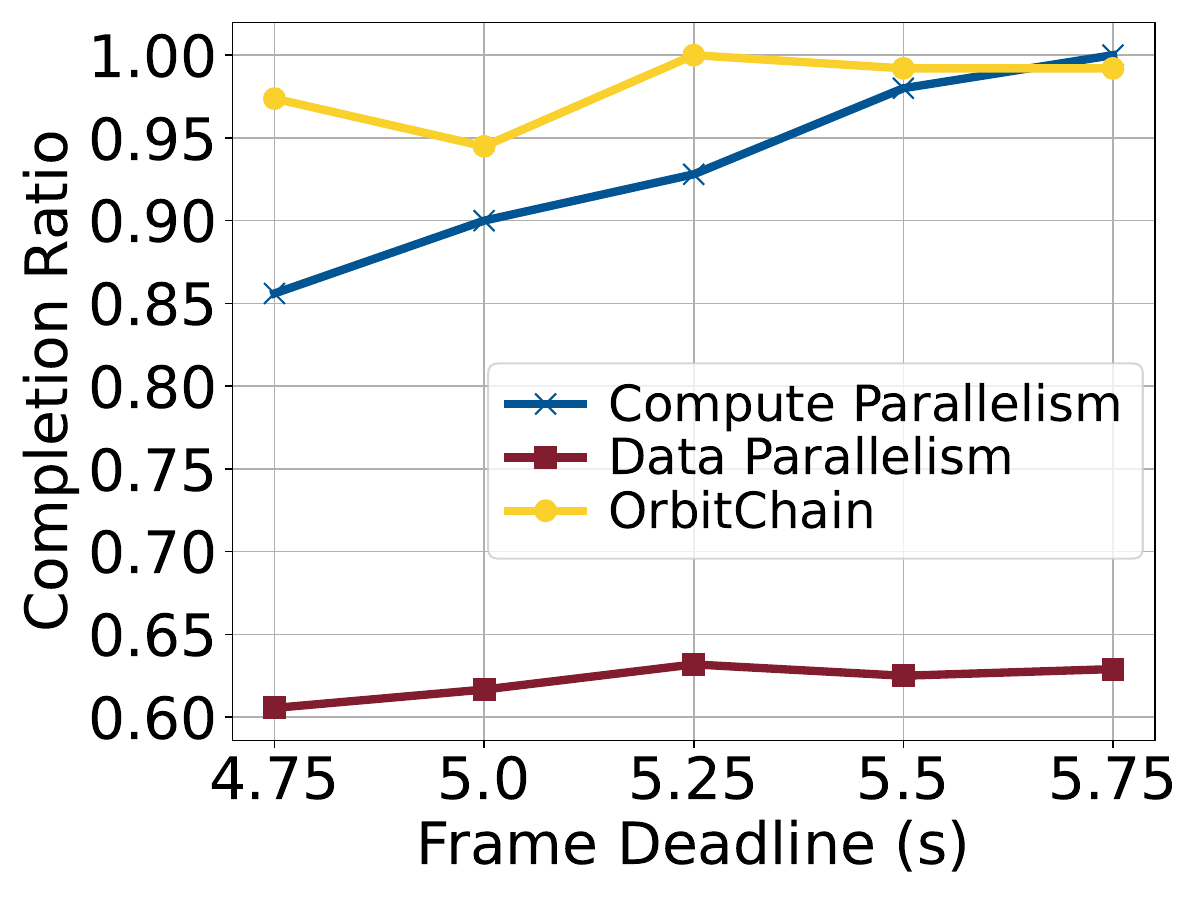}
\label{fig:completion_ratio:jetson_cloud_object_water}}
\hfil
\subfloat[Cloud{--}land use{--}crop\&water.]{\includegraphics[width=0.23\textwidth]{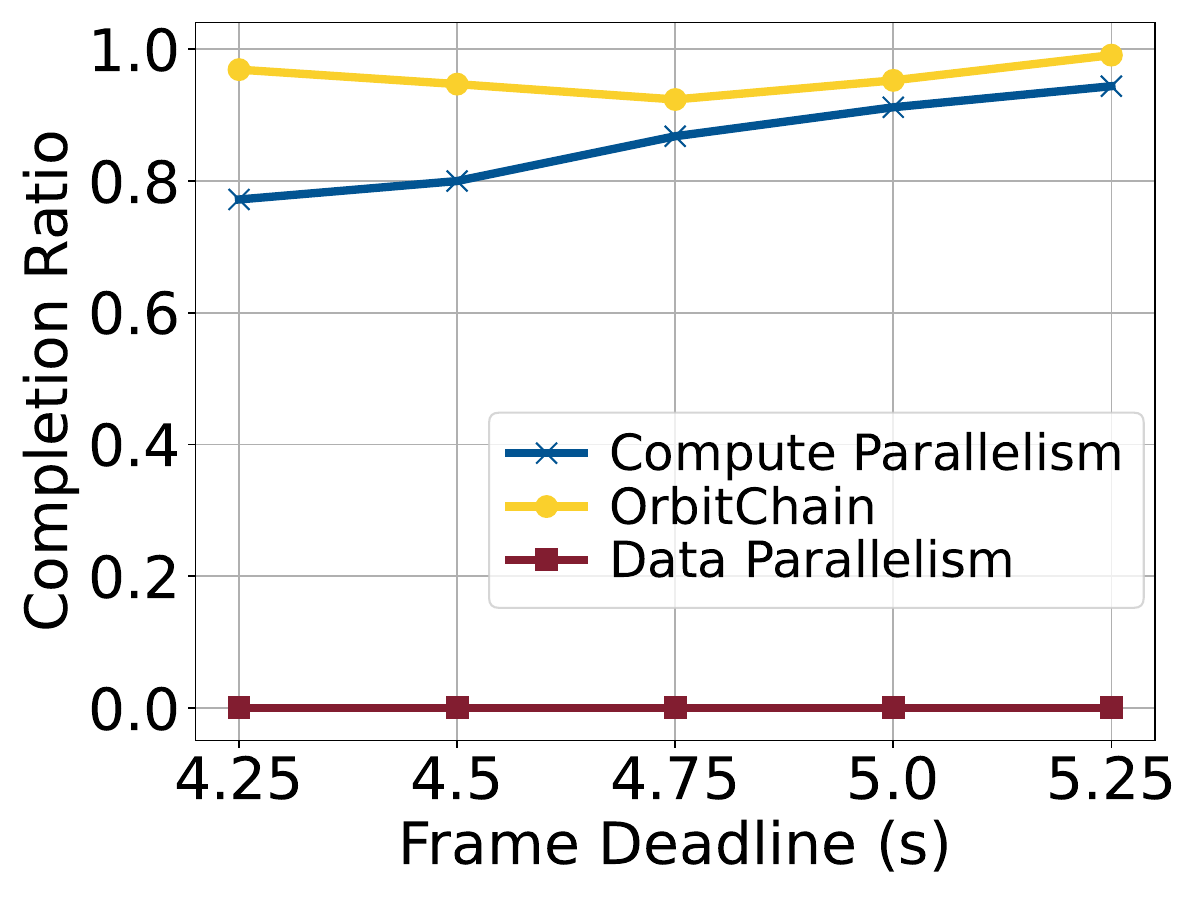}
\label{fig:completion_ratio:pi}}
\hfil
\vspace{-1em}
\caption{Analytics task completion ratio on NVIDIA Jetson Orin Nanos.}
\label{fig:completion_ratio}
\vspace{-1em}
\end{figure*}

\subsection{Experiment Setup}
\noindent\textbf{Dataset.}
We use the LandSat8 Cloud Cover dataset~\cite{foga2017cloud,zhu2018cloud} for evaluation. 
We extract the RGB bands and divide each frame into $640$pixel $\times$ $640$pixel tiles. 
We evaluate both chain-like and span-like OEC workflows on NVIDIA Jetson Orin Nanos, as well as different model compositions on both Jetsons and Raspberry Pis. 
The randomness in evaluation results are averaged out over running each experiment with 10 distinct frames, which include 250 to 1000 tiles in total.

\textbf{Analytics functions.}
We evaluate the four functions in the workflow as in Fig.~\ref{fig:application}.
The Sentinel-2 Cloud Mask~\cite{aybar2022cloudsen12}, Eurosat~\cite{helber2018introducing,helber2019eurosat}, and  Satellite Images of Water Bodies~\cite{waterbody_dataset} datasets are used for training cloud detection, land use classification, crop monitoring, and water monitoring  respectively.
On Raspberry Pis, we deployed four YOLO-based analytics functions. 
On Jetsons, we deployed models with different architectures:
cloud detection based on MobileNetV2~\cite{sandler2018mobilenetv2}, water monitoring based on EfficientNet~\cite{tan2019efficientnet}, and land use classification and crop monitoring based on YOLOv8n~\cite{ultralytics}.

\textbf{Parameters.}
Discount coefficients for CPU capacity $\alpha$ and GPU time $\beta$ are set as $0.95$ for Jetson and $0.9$ for Raspberry Pi.
The frame deadlines ranges from $4.75$ to $5.5$ seconds for Jetson and $12$ to $16$ seconds for Raspberry Pi, both matching existing constellations~\cite{2016sentinel2}.
Each frame is divided into $100$ tiles for Jetson and $25$ tiles for Raspberry Pi.
The inter-satellite interval for revisiting the same ground location is $10$ seconds for Jetson and $15$ seconds for Raspberry Pi.
Workload distribution ratios are by default $0.5$.
To model orbit shift, we consider two subsets including the first and the first two satellites in a constellation, with $5$ and $20$ unique images respectively.
Available power on each satellite is $7$ Watts~\cite{3u}.

\textbf{Compared frameworks.}
We compare {\oursol} with the two baseline frameworks: data parallelism and compute parallelism.
In \textbf{\textit{data parallelism}}~\cite{denby2020orbital}, each satellite hosts all analytics functions, and the workload is evenly distributed across all satellites.
In \textbf{\textit{compute parallelism}}, functions are deployed as one pipeline, sequentially across the constellation while balancing each satellite's overall workload. %
When evaluating workload routing specifically, we also compare {\oursol} to a workload balancing variant (\textbf{\textit{load spraying}}), where each function instance sprays workload among downstream instances inspired by network load balancing~\cite{song2023network}.

\textbf{Metrics.}
We evaluate four metrics:
\textit{(1) Completion ratio:} number of analyzed tiles over received ones for each analytics function, averaged over all functions in a workflow. 
\textit{(2) Inter-satellite traffic:} total data volume transmitted over all inter-satellite links per frame.
\textit{(3) Analyzable tiles:} number of tiles per frame that can be analyzed by entire workflow as sensors with higher resolution or coverage are equipped.
\textit{(4) End-to-end latency:} time to analyze all tiles in a frame is measured from when the first function begins analyzing the first tile, till the last function completes the last tile.

\begin{figure*}[th]
\centering
\subfloat[Cloud{--}land use{--}crop.]{\includegraphics[width=0.23\textwidth]{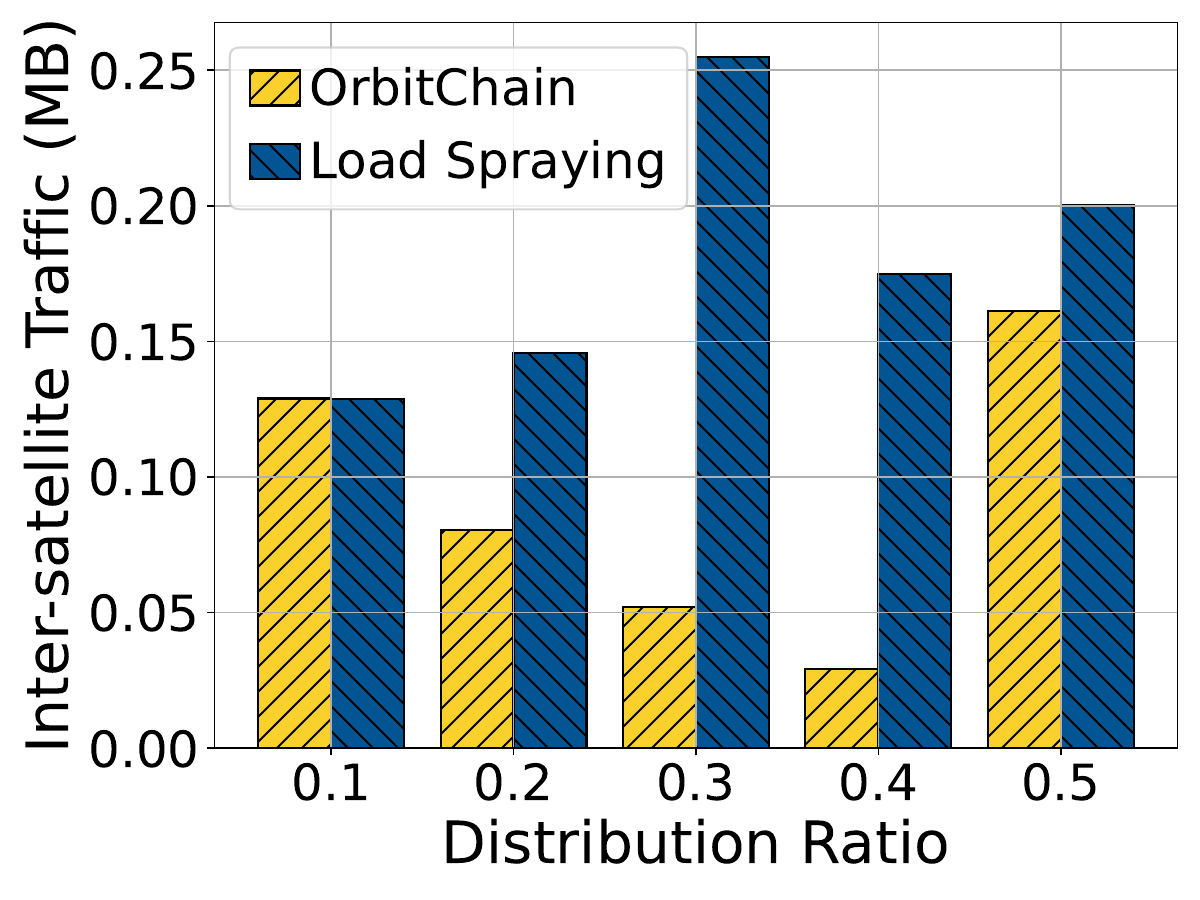}
\label{fig:comm_overhead:jetson_cloud_landuse_object}}
\hfil
\subfloat[Cloud{--}land use{--}water.]{\includegraphics[width=0.23\textwidth]{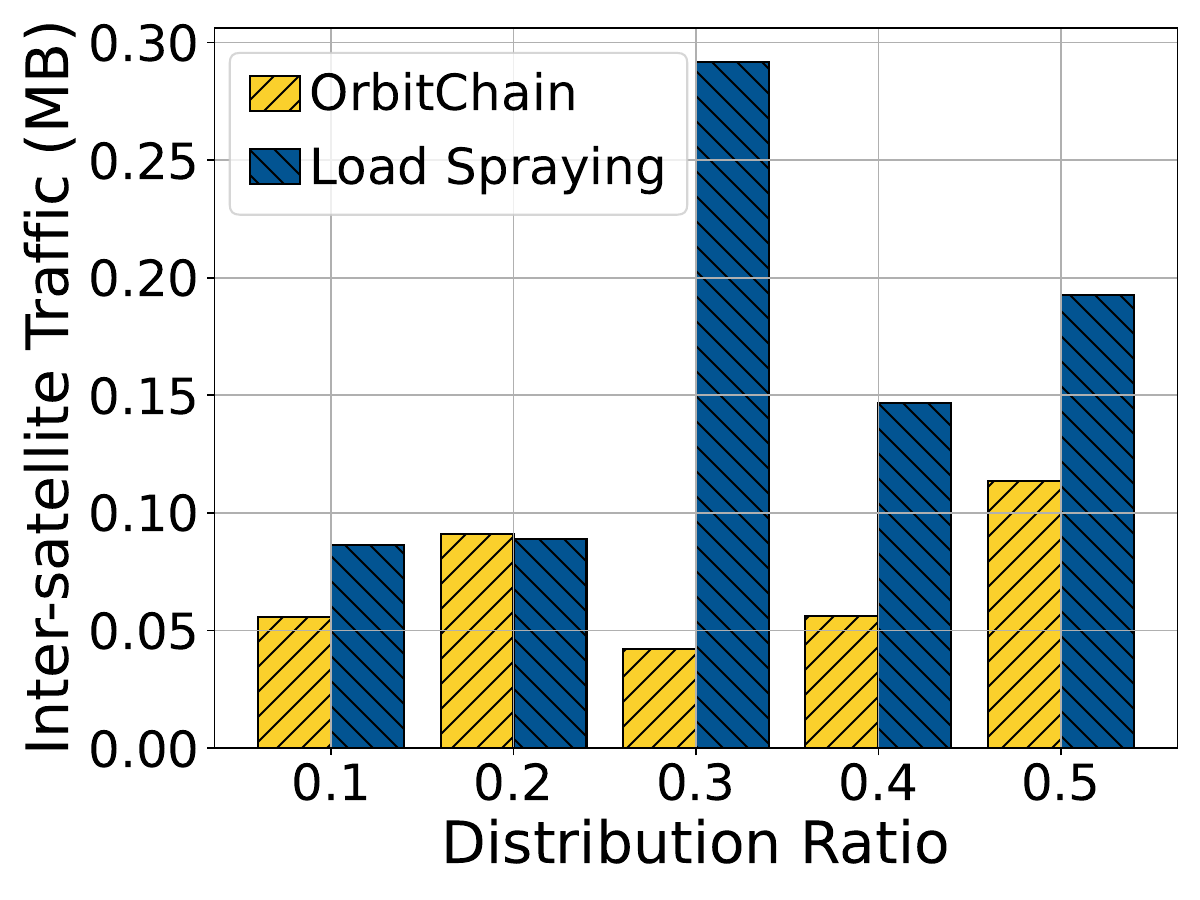}
\label{fig:comm_overhead:jetson_cloud_landuse_water}}
\hfil
\subfloat[Cloud{--}crop and water.]{\includegraphics[width=0.23\textwidth]{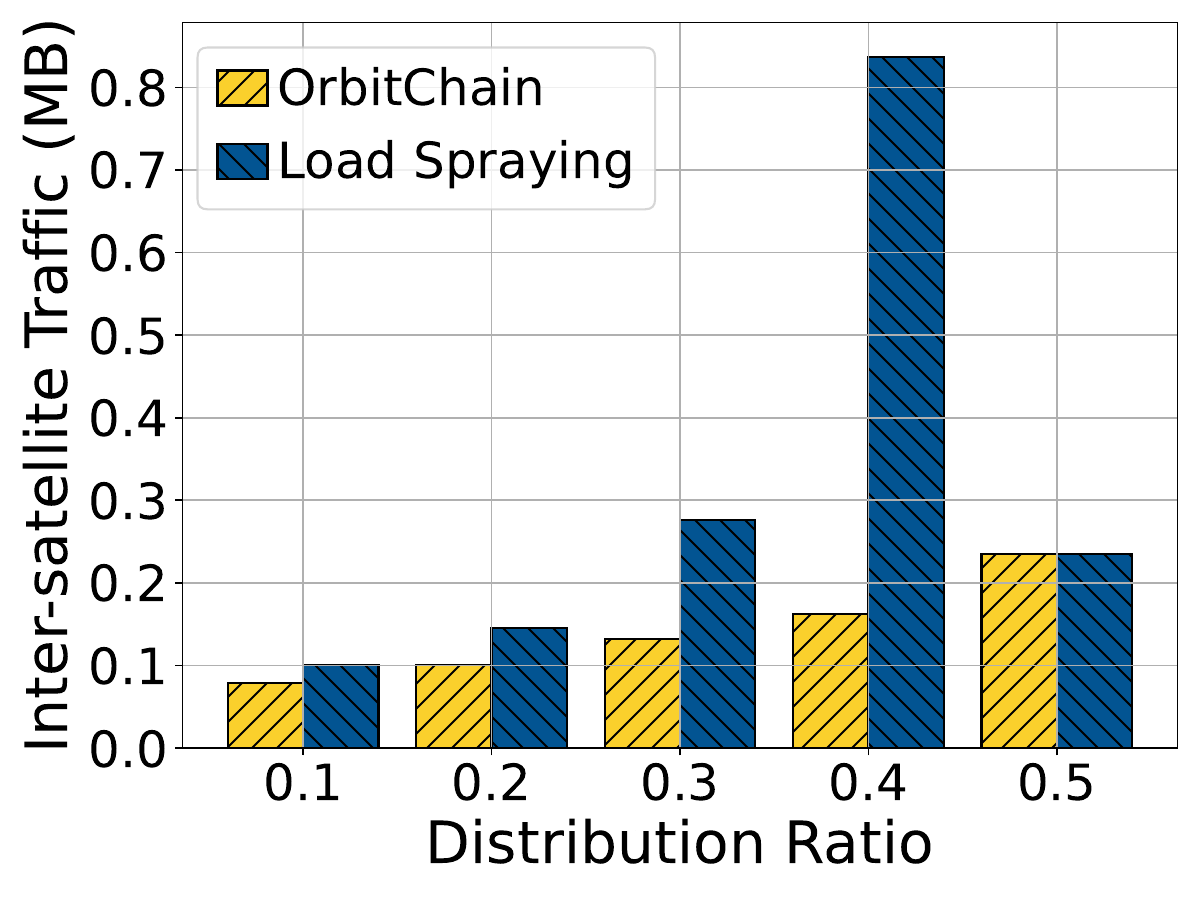}
\label{fig:comm_overhead:jetson_cloud_object_water}}
\hfil
\subfloat[Cloud{--}land use{--}crop\&water.]{\includegraphics[width=0.23\textwidth]{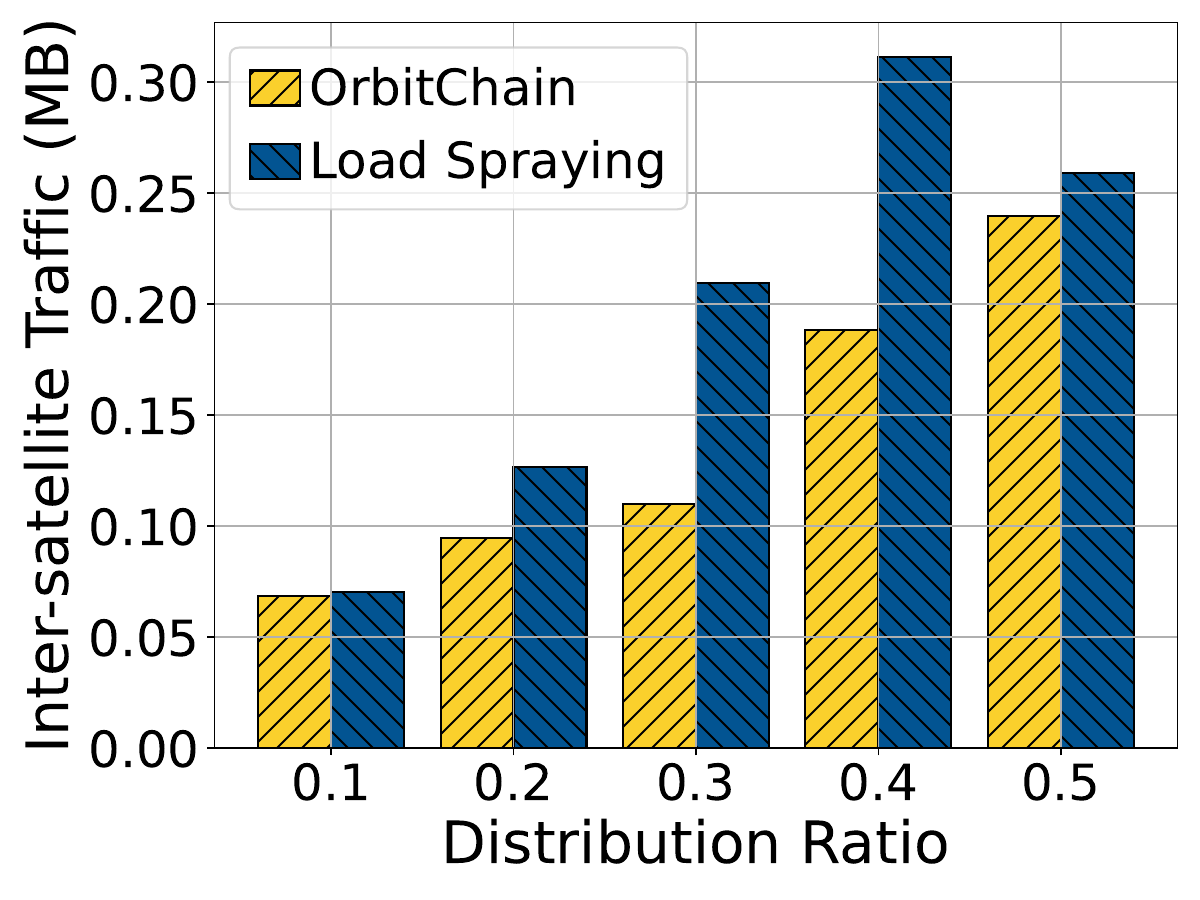}
\label{fig:comm_overhead:pi}}
\hfil
\vspace{-1em}
\caption{Average per-frame inter-satellite communication overhead on NVIDIA Jetson Orin Nanos.}
\label{fig:comm_overhead}
\vspace{-1em}
\end{figure*}

\begin{figure}[t]
\centering
\subfloat[Completion ratio.]{\includegraphics[width=0.232\textwidth]{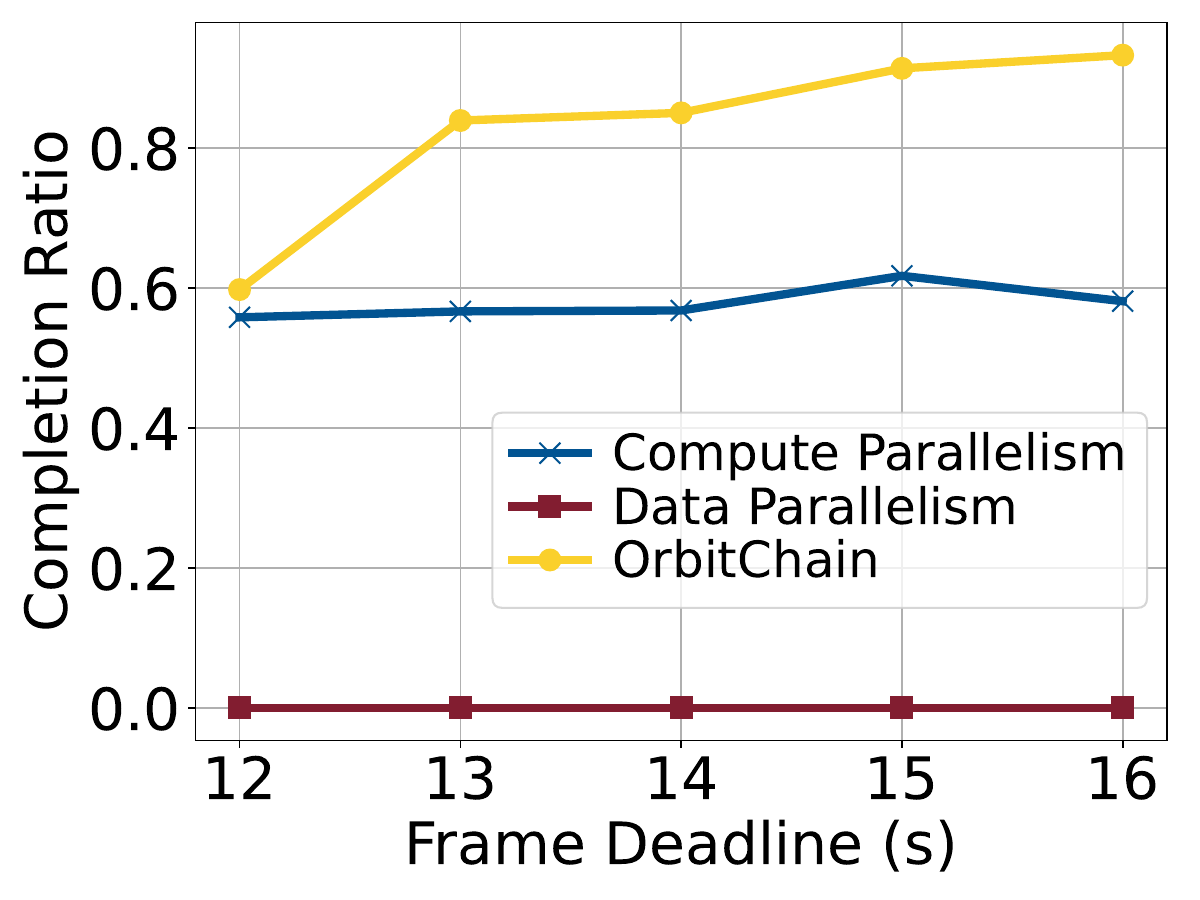}
\label{fig:raspberry_pi:completion_ratio}}
\hfil
\subfloat[Communication overhead.]{\includegraphics[width=0.232\textwidth]{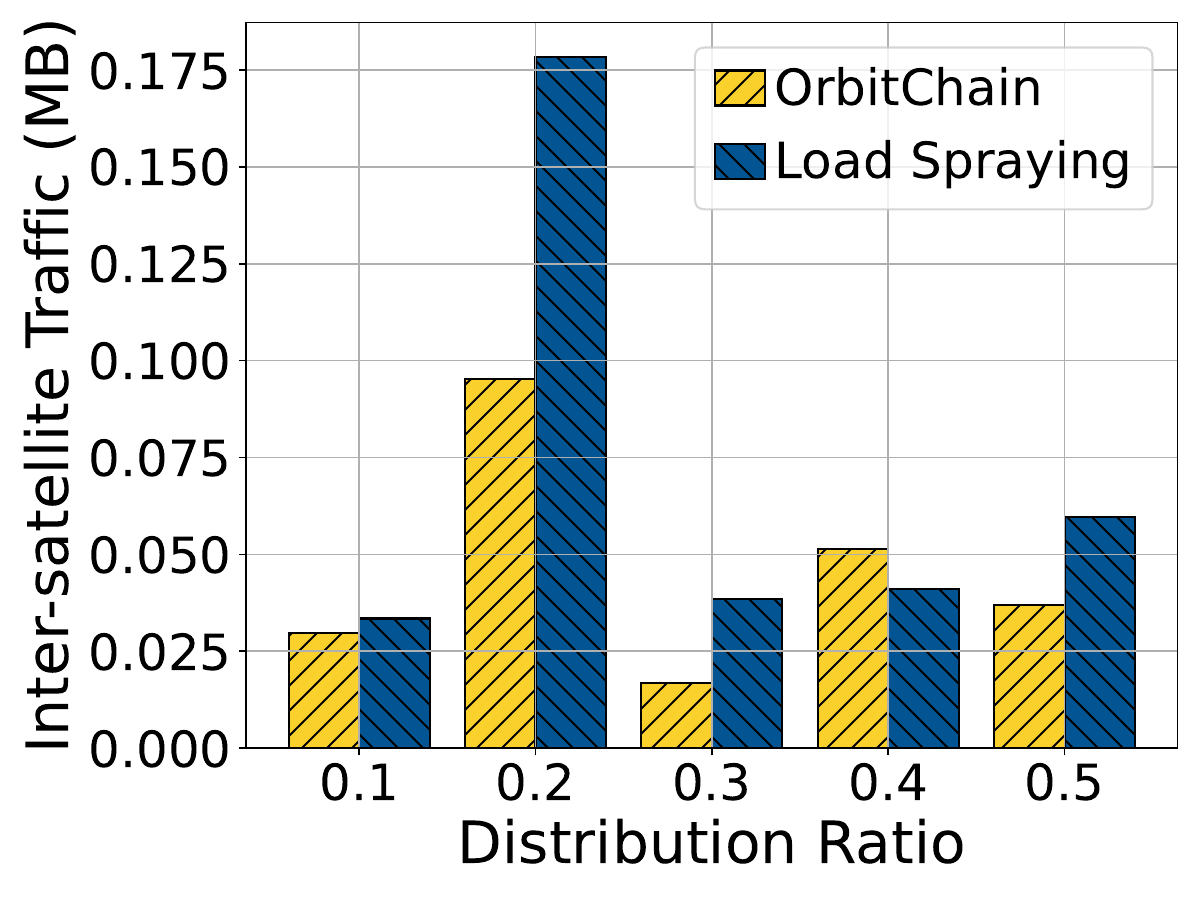}
\label{fig:raspberry_pi:communication}}
\hfil
\vspace{-1em}
\caption{Analytics task completion ratio and communication overhead on Raspberry Pis.}
\label{fig:raspberry_pi}
\end{figure}

\subsection{Evaluation Results}

\noindent
Below, we present our evaluation results on the Jetson and Raspberry Pi devices while comparing to the baselines.

\textbf{(1) Analytics task completion.}
Fig.~\ref{fig:completion_ratio} shows the completion ratio of compared frameworks on Jetson.
{\oursol} consistently achieves nearly $100\%$ completion ratio across different frame deadlines, while both baselines achieve lower completion leading to buffer buildup.
The superior performance stems from {\oursol}'s flexible orchestration, maximizing the utility of on-board computing.
For data parallelism, the low completion ratio comes from resource contention on each satellite as in Fig.~\ref{fig:data_parallel}\subref{fig:data_parallel:resource_contention}, 
limiting the total capacity of all functions.
Especially with all four functions, Jetson cannot host all four functions, and hence data parallelism fails to instantiate the workflow leading to $0\%$ completion.
In Fig.~\ref{fig:raspberry_pi}\subref{fig:raspberry_pi:completion_ratio}, we observe the same trends on Raspberry Pi.
Data parallelism cannot instantiate all four functions even with virtual memory enabled.
Meanwhile, at a $16$-second frame deadline, {\oursol} achieves $60\%$ higher completion ratio than compute parallelism.
The performance of compute parallelism improves with longer frame deadline on Jetsons but not on Raspberry Pis.
As GPU acceleration achieves over $10\times$ faster tile processing than CPU-only execution, it benefits much more from increased processing time limit.

\textbf{(2) Communication overhead.}
To analyze the impact of workload routing, we compare the communication overhead per frame between {\oursol} and load spraying, the latter of which ignores inter-satellite communications when routing workloads.
In Fig.~\ref{fig:comm_overhead}, we vary the distribution ratio on cloud detection, and notice that in both platforms, {\oursol} achieves significantly less communication overhead.
On average, {\oursol} saves up to $45\%$ inter-satellite traffic on Jetsons and $25\%$ on Raspberry Pis compared to communication-agnostic workload scheduling.
Also, as we combine the results with Fig.~\ref{fig:profiling_props}\subref{fig:profiling_props:communication}, both {\oursol} and load spraying reduce communication overhead by orders of magnitude compared to sending raw data, demonstrating the strong advantage of utilizing local sensing functions to replace heavy inter-satellite communications.

\begin{figure}[t]
\centering
\subfloat[Jetson Orin Nano]{\includegraphics[width=0.232\textwidth]{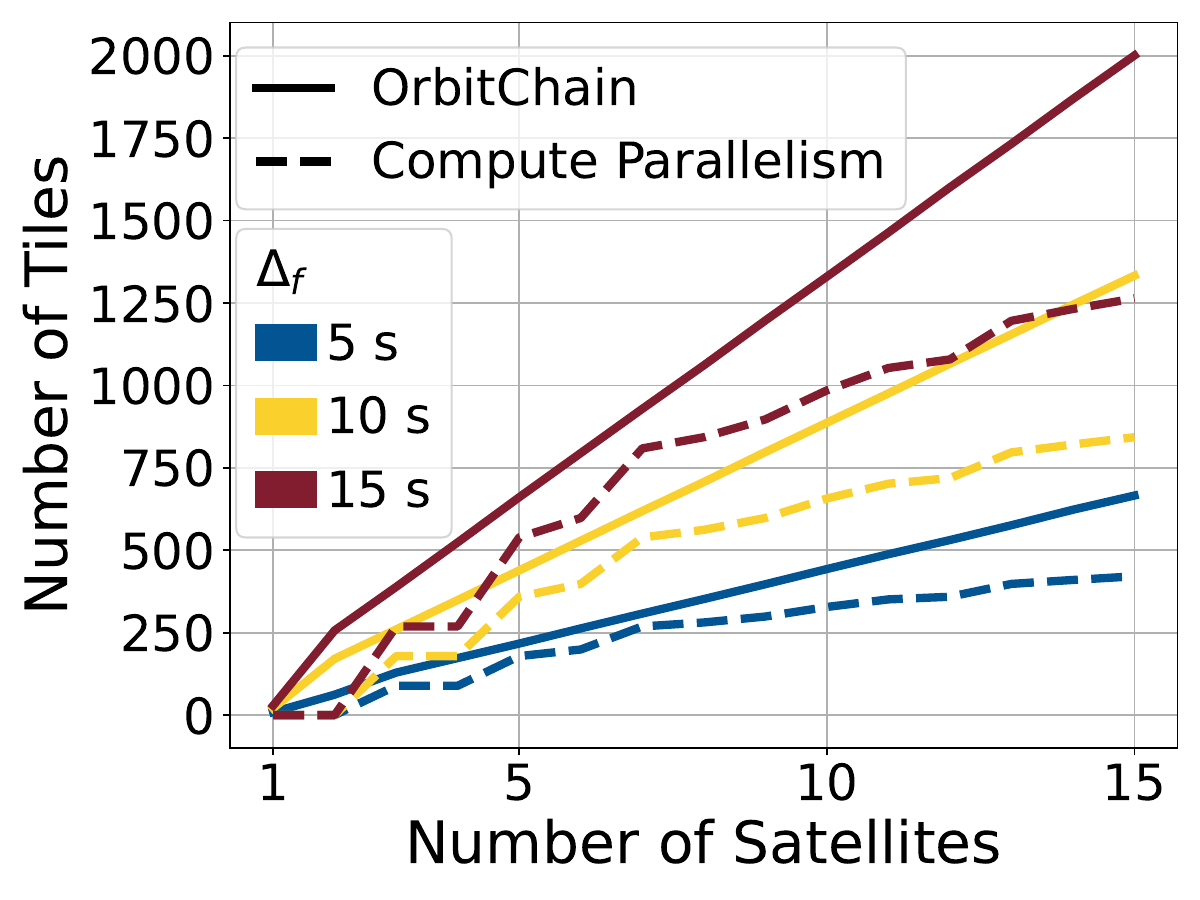}
\label{fig:frame_size:jetson}}
\hfil
\subfloat[Raspberry Pi]{\includegraphics[width=0.232\textwidth]{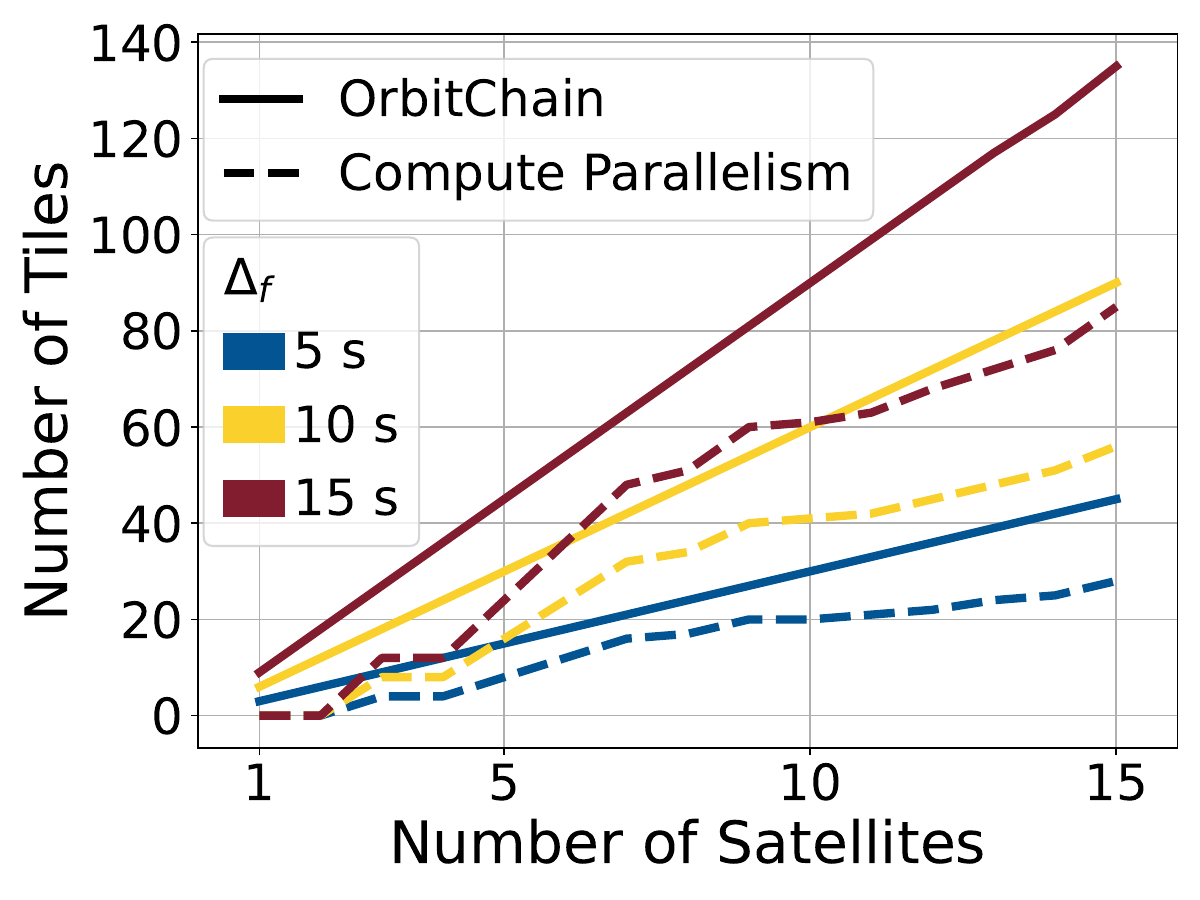}
\label{fig:frame_size:pi}}
\hfil
\vspace{-1em}
\caption{Number of analyzable tiles within $\Delta_f$.}
\label{fig:frame_size}
\end{figure}

\textbf{(3 )Analyzable tiles.}
We next evaluate how {\oursol} and compute parallelism scale to analyze more tiles as the resolution or coverage of on-board sensors increases.
Due to hardware availability, we evaluate this by using the formulation in Section~\ref{sec:framework:deployment} to check if there is a feasible deployment that can analyze a given number of tiles, using an increasing number of Jetson or Raspberry Pi satellites based on the profiled data.
We evaluate the full workflow in Fig.~\ref{fig:application} for Jetsons and Raspberry Pis.
As in Fig.~\ref{fig:frame_size}, the number of tiles that {\oursol} can analyze is, on average, $42\%$ and $71\%$ higher than compute parallelism for Jetsons and Raspberry Pis, respectively. 
This is because as the number of satellites increase, {\oursol} can even more flexibly utilize cross-satellite resources, and its overall capacity scales linearly as the number of satellites.
Also, as we increase the frame deadline, both frameworks can analyze more tiles, though the comparison between them remains.

\textbf{(4) End-to-end analytics latency.}
Our final goal is to demonstrate real-time analytics performance of {\oursol}.
We evaluate a \texttt{cloud--land use--crop} workflow for Jetsons and the full workflow for Raspberry Pis.
We use \texttt{tc} to emulate channel bandwidth, and record the end-to-end latency for each tile; detailed channel simulation is shown in Appendix~\ref{appdx:intersatellite_links}.
We further breakdown end-to-end latency into \emph{processing delay}, \emph{communication delay} and \emph{revisit delay}.
Revisit delay is incurred when a downstream instance waits for the satellite to move over the target area for capture, and is accumulated across multiple satellites in a pipeline but in parallel to the processing and communication delays.
The frame's latency is taken as maximum end-to-end latency of any tile.

\begin{figure}[t]
\centering
\subfloat[Jetson Orin Nano.]{\includegraphics[width=0.232\textwidth]{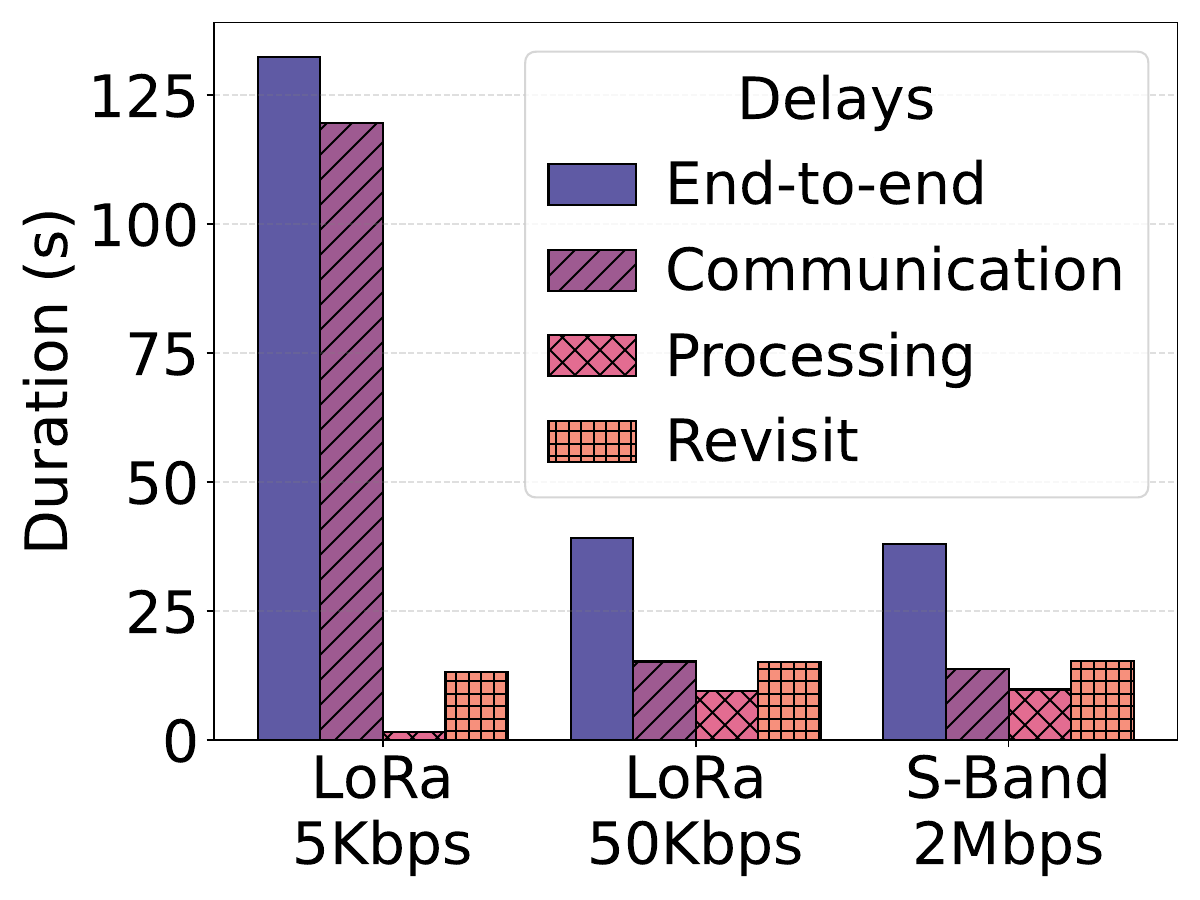}
\label{fig:bandwith:jetson}}
\hfil
\subfloat[Raspberry Pi.]{\includegraphics[width=0.232\textwidth]{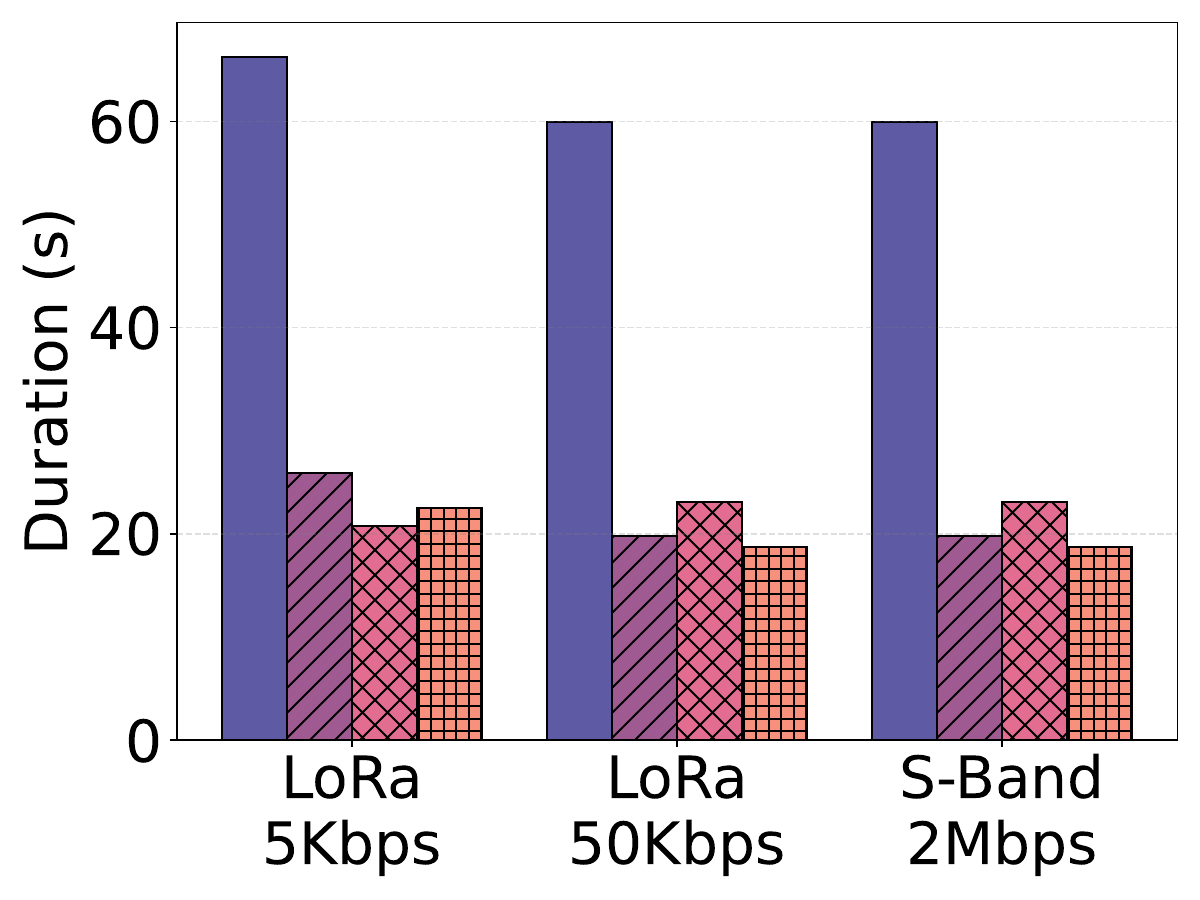}
\label{fig:bandwith:pi}}
\hfil
\vspace{-1em}
\caption{Bandwidth v.s. analysis latency.}
\label{fig:bandwith}
\end{figure}

As shown in Fig.~\ref{fig:bandwith}\subref{fig:bandwith:jetson}, using a 5 Kbps LoRa channel, an $100$-tile frame completes in less than 3 minutes on a 3-satellite Jetson constellation.
When bandwidth increases to 50 Kbps for high-speed LoRa, end-to-end latency drops to under 30 seconds, at which point the inter-satellite channel is no longer the bottleneck.
Meanwhile, inter-satellite bandwidth has limited impact on the end-to-end latency on the Raspberry Pi constellation, which is dominated by processing delay.
Note that this end-to-end latency does not violate the frame deadline constraint, as that constraint is on a per-function basis and only serves to avoid buffer buildup at each function.
Overall, the analytics latency is \emph{orders of magnitude lower than current constellations} which is in the order of hours to days, and is \emph{sufficiently low for most real-time Earth analytics applications} such as fire or flood monitoring.

Looking at the breakdown, we observe that no single source of delay is a major bottleneck, including the revisit delay.
The processing delay can be reduced by better hardware and/or energy input.
Communication delay is reasonably low even with low-cost low-power inter-satellite links such as LoRa, which is readily available on existing satellites.
However, if raw data is transmitted instead of intermediate results, communication delay (and hence also power consumption for communications) would increase by orders of magnitude as in Fig.~\ref{fig:profiling_props}\subref{fig:profiling_props:communication}, in which case high-power high-speed optical links could be the only option for in-orbit multi-satellite collaborative analytics.
This again showcases {\oursol}'s key advantage in utilizing data locality for in-orbit analytics.%

\section{Conclusion}
\label{sec:conclusion}
In this paper, we proposed {\oursol}, a framework for real-time in-orbit Earth observation analytics tasks.
{\oursol} orchestrates computational resources across multiple satellites in an Earth observation constellation to support complex real-time analytics tasks.
To mitigate inter-satellite communication overhead, {\oursol} uniquely utilizes data locality enabled by overlapping satellite views, removing the need to transmit raw data across satellites.
It employs intelligent function placement, resource allocation and proactive workload routing to efficiently utilize constellation-wide computing and energy resources, while accounting for misaligned satellite coverage due to natural orbit shifts.
Experiments on a hardware-in-the-loop OEC testbed have validated that {\oursol} achieves superior analytics throughput compared to existing framework, conserves energy and inter-satellite bandwidth, while delivering Earth analytics insights in minutes to support critical time-sensitive applications.

\bibliographystyle{ACM-Reference-Format}

\appendix
\section*{Appendices}

\section{Orbital Edge Testbed}
\label{appdx:testbed}
\begin{figure}[ht]
\centering
\includegraphics[width=0.45\textwidth]{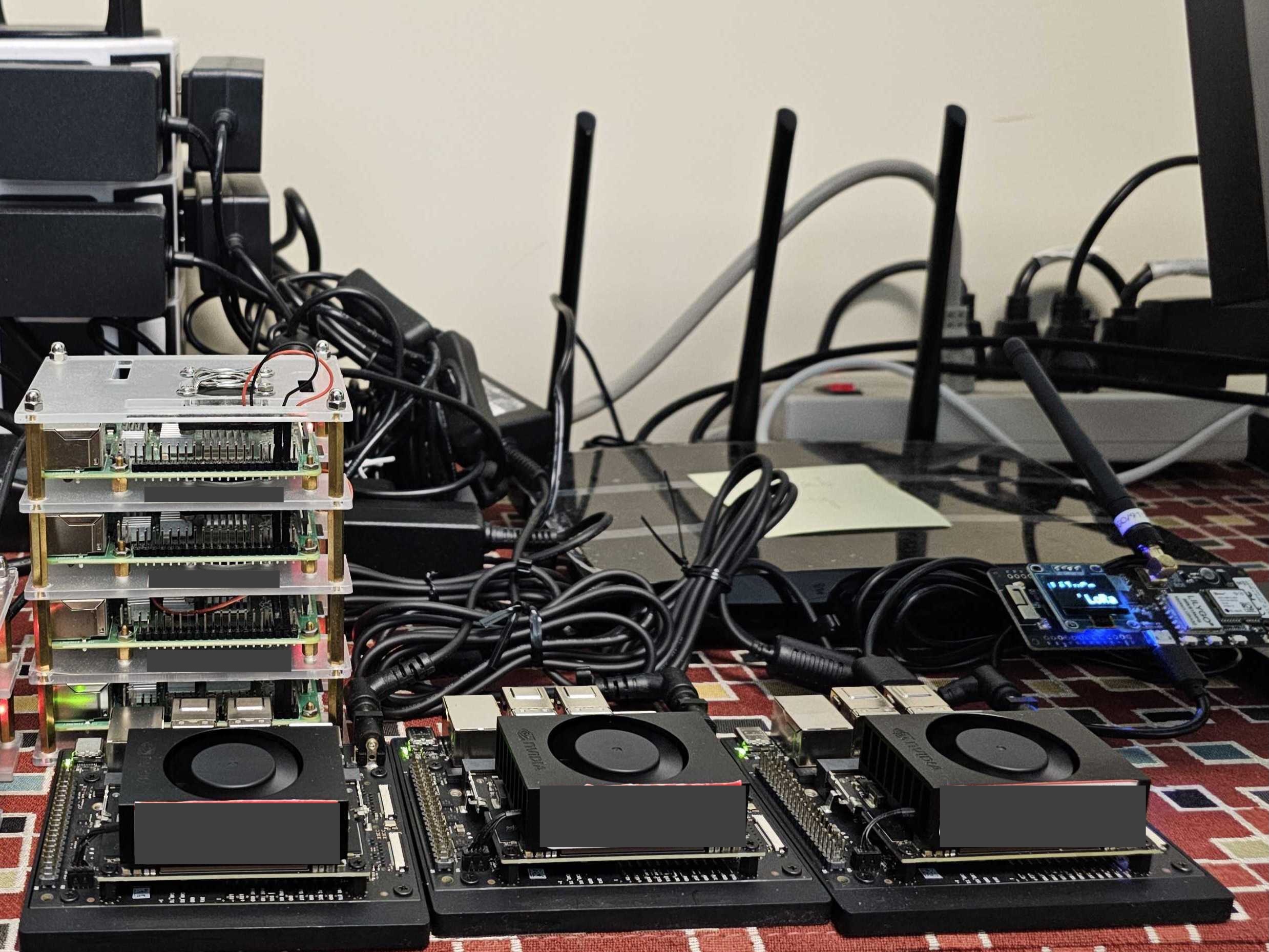}
\caption{Orbital edge testbed.}
\label{fig:testbed}
\end{figure}

Our orbital edge testbed is presented in Fig.~\ref{fig:testbed}.
In this section, we introduce its implementation in detail.

\noindent\textbf{Hardware devices.}
Our orbital edge testbed consists of three Nvidia Jetson Orin Nanos and four Raspberry Pis.
These edge devices are interconnected through a programmable OpenWRT wireless access point, whose connection can be controlled with tools like \texttt{tc}.
Each Jetson Orin Nano features $8$GB of integrated memory shared between RAM and VRAM, along with an NVIDIA Ampere GPU containing $32$ Tensor Cores.
In $7$W power mode, the device activates four Arm Cortex®-A78AE v8.2 64-bit CPUs operating at a base frequency of $729$MHz when idle and at maximum for $1.7$GHz.
Each Raspberry Pi features a Quad-core Cortex-A72 (ARM v8) 64-bit SoC running at a maximum of $1.8$GHz, and has 4GB of RAM.
Two desktops (not shown in the figure) are connected to these edge devices to control experiments and collect data, but not participate in the computation.

\noindent\textbf{Operating system and onboard services.}
The Nvidia Jetson Orin Nanos run Jetpack 5.1, an Ubuntu-based operating system, operating without a graphical interface to conserve computational resources.
The Raspberry Pis also run the Ubuntu 22.04 operating system in headless mode.
Using Docker, all computations run in containerized services that enable flexible deployment, resource allocation, and metric collection.
Container management can be performed remotely from the connected desktops, minimizing interference with experimental workloads.

\noindent\textbf{Monitoring and tracing.}
We use node exporter on each device to collect utilization metrics every second.
The node exporter's CPU usage is negligible, consuming only $0.3\%$ CPU cores and $14$MB memory.
A Prometheus monitoring service host on the desktop collects utilization data from the orbital edge devices every 10 seconds.
We maintain a Springboot-based image server for image downloading and serving.
For efficiency, we pre-load all data to edge devices before experiments.
Network traces can be collected by instrumenting the computation services with frameworks like OpenTelemetry and exporting network traces to the Jaeger collector host on the desktop server.

\section{Limitations of Ground-assisted Earth Observation.}
\label{appdx:ground_assist_framework}
\noindent
In traditional Earth observation workflows, satellites capture remote sensing data and download it to the ground for analysis.
This approach leads to long delivery delays due to limited downlink channel capacity.
To address this issue, some OEC satellites perform lightweight in-orbit analysis to filter out low-value data and only keep high-value data for download, thus using the downlink channel more efficiently~\cite{tao2024known,denby2023kodan}.
In this case study, we first evaluate the time intervals between consecutive satellite-ground connections to demonstrate why real-time Earth observation analytics is infeasible when ground stations participate in data analysis.
We then compare the downlink channel capacity with the generated data volume and notice that in-orbit image filtering is not capable to make all the high-value data being timely analyzed.

\begin{figure}[ht]
\centering
\subfloat[CDF of satellite-ground connection intervals.]{\includegraphics[width=0.22\textwidth]{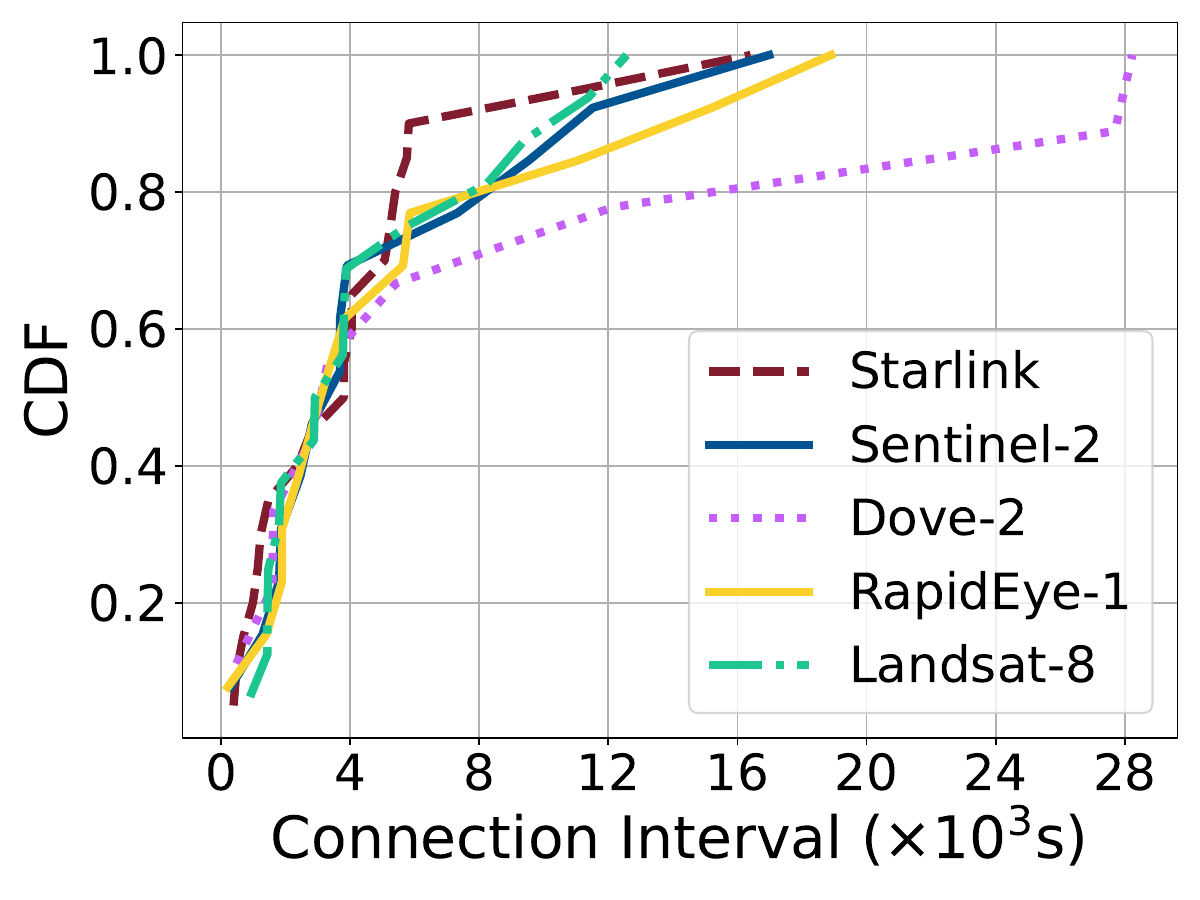}
\label{fig:gs_conn:conn_interval}}
\hfil
\subfloat[Downlinkable data ratio of the previous interval.]{\includegraphics[width=0.22\textwidth]{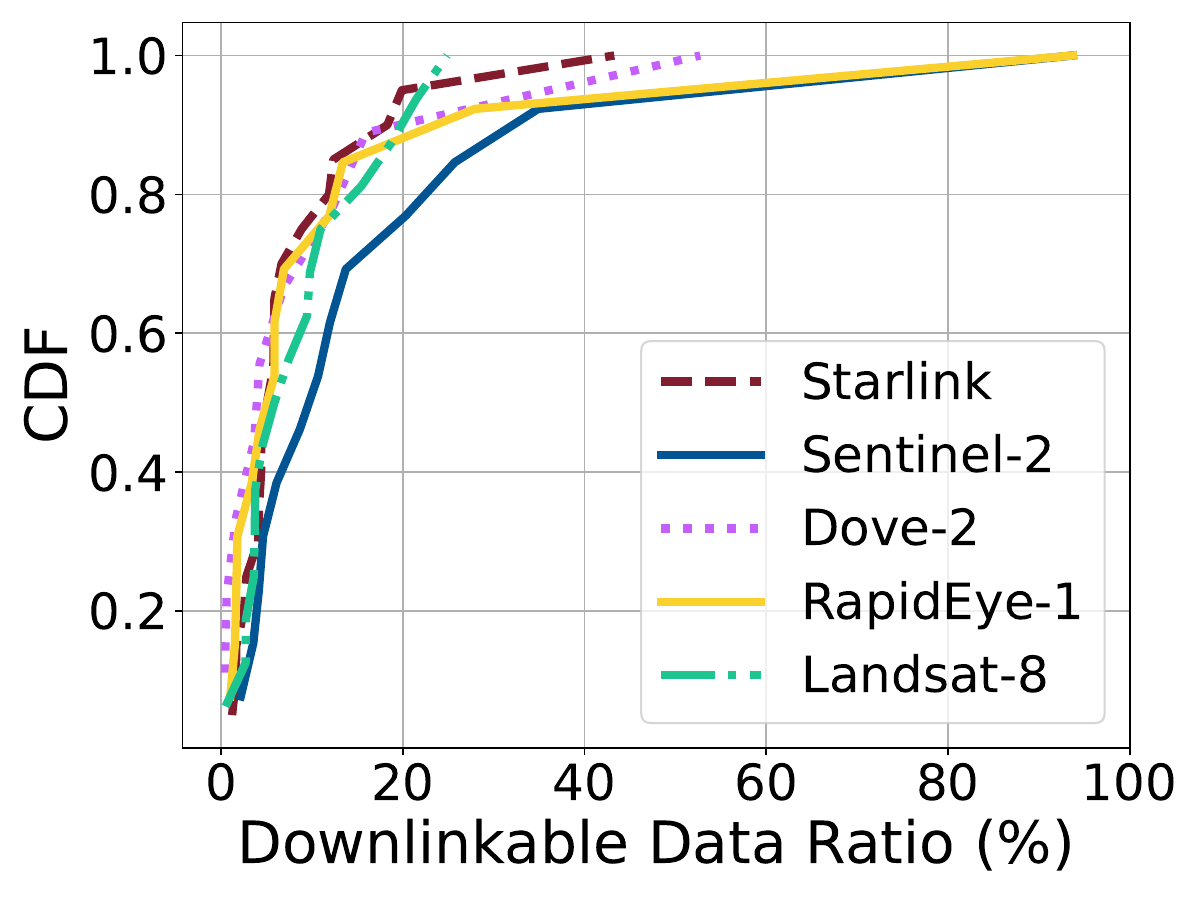}
\label{fig:gs_conn:downlinkable_data_ratio}}
\hfil
\caption{(a) Satellite-ground connection intervals. (b)) Downlinkable data ratio in each satellite-ground connection. $50\%$ of data in the previous interval has already been filtered out with single-satellite OEC.}
\label{fig:gs_conn}
\end{figure}

\noindent\textbf{Experiment setup.}
We simulate the orbits of five mainstream LEO constellations (Starlink, Sentinel-2, Dove-2, RapidEye, and Landsat-8) over a $24$-hour period with the Hypatia LEO simulator~\cite{hypatia}.
We track both the consecutive connection intervals and the duration of each connection. 
The simulation uses $10$ ground stations in the most populated areas. 
We selected the most populations because ground stations with network and computational resources are typically located near population centers.

\noindent\textbf{Ground-satellite connection interval.}
We first present the cumulative distribution of time intervals between satellite-ground connections in Fig.~\ref{fig:gs_conn}\subref{fig:gs_conn:conn_interval}. The analysis reveals that in more than half of cases, satellites must wait at least one hour to connect with the next ground station for data download.
This finding demonstrates that real-time analysis through ground stations is not feasible for time-sensitive analytics tasks requiring minute-level responses.

\noindent\textbf{Downloadable data ratio.}
We further analyze the data download capacity during intermittent connections, with results shown in Fig.~\ref{fig:gs_conn}\subref{fig:gs_conn:downlinkable_data_ratio}. 
We convert ground-track length to data volume, where a $110 \times 110$Km area generates a $500$MB data frame, using the Sentinel-2 constellation as a reference,~\cite{2016sentinel2}. 
The results demonstrate that even when filtering out $50\%$ of in-orbit data through single-satellite OEC, none of the mainstream constellations can fully download their in-orbit data.
Hence, we have the following observation:
\begin{observation}
\label{obsv:ground_assisted_analytics}
With current ground satellite infrastructure, Earth observation analytics cannot be completed in real-time, nor can they process all captured in-orbit data, when relying on assistance from ground stations for analysis.
\end{observation}

\section{Inter-satellite Links Simulation}
\label{appdx:intersatellite_links}
\begin{figure}[t]
\centering
\includegraphics[width=0.37\textwidth]{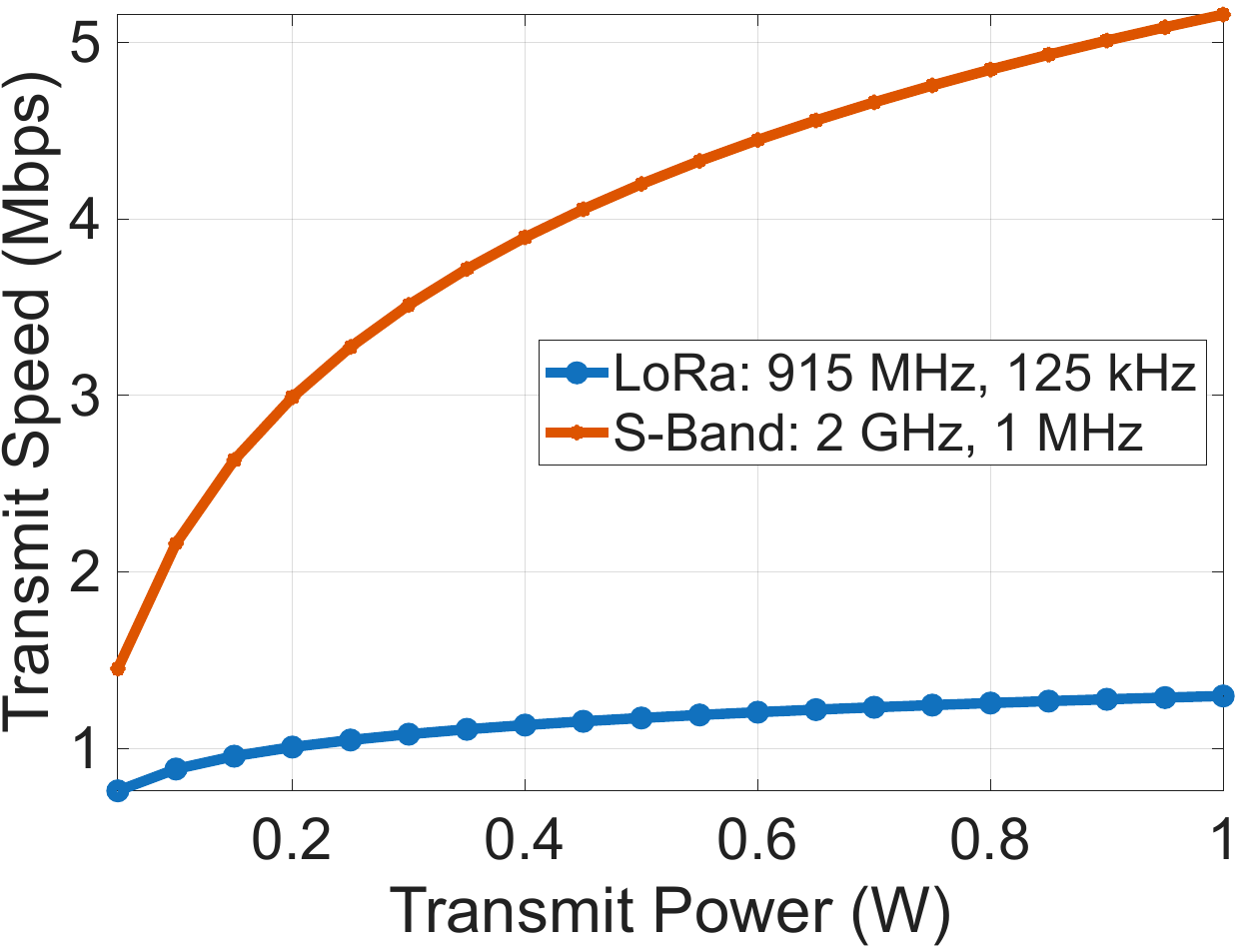}
\caption{Simulation results for transmit power v.s. transmit speed of two common inter-satellite links S-Band and LoRa.}
\label{fig:isl_simulation}
\end{figure}
\noindent\textbf{Inter-satellite geometry.}
We consider a dense same-orbit LEO constellation in which adjacent satellites pass the same ground location with a temporal separation of approximately $5$~s. 
For a representative LEO orbit with a period of about $90$~min, this temporal offset corresponds to an along-track angular separation of $\Delta\theta \approx 2\pi \Delta t / T$, resulting in an inter-satellite distance of roughly $40$--$50$~km.
Such a short-range inter-satellite geometry significantly reduces free-space path loss and relaxes antenna pointing requirements compared to conventional cross-orbit inter-satellite links spanning hundreds or thousands of kilometers.

\noindent\textbf{Communication model.}
We considered two inter-satellite communication technologies: a LoRa sub-GHz narrowband link and a conventional S-Band link.
The LoRa-like configuration operates at $915$~MHz with $125$~kHz--$1$~MHz bandwidth and low-gain ($2$~dBi) quasi-omnidirectional antennas, reflecting typical LoRa designs that emphasize ultra-low power consumption, high robustness, and relaxed pointing requirements at the expense of spectral efficiency~\cite{adelantado2017understanding}.
In contrast, the S-Band link operates in the $2.2$--$2.4$~GHz range with MHz-level bandwidth ($1$--$2$~MHz), consistent with widely adopted CubeSat and small-satellite communication systems supporting Mbps-level data rates~\cite{selva2012survey,ccsds401}.

\noindent\textbf{Simulation results.}
Figure~\ref{fig:isl_simulation} shows the achievable inter-satellite throughput as a function of transmit power for LoRa and S-Band channels under the same short-range, same-orbit geometry, with an inter-satellite separation of approximately 40–50~km.
The figure shows a monotonic increase in throughput with transmit power for both channels, while revealing a clear separation between the achievable rate regions of LoRa and S-Band links.
S-Band can reach approximately $2$~Mbps transmit speed with less than $0.1$~W power consumption, making it suitable as a duty-cycled communication channel for inter-satellite data delivery. LoRa stays under $1.5$~Mbps across different power levels but can provide broader, longer-range communication without the need for point-to-point alignment. It can also be remained always-on for constellation-to-constellation result delivery and satellite-ground result offloading.
Bands with higher frequencies, though they provide higher speed, usually require more accurate positioning and inter-satellite alignment, and may incur high costs in satellite posture correction.
Meanwhile, constellations with larger inter-satellite distances further reduce transmit speed for a given transmit power.
Combined with the noisy radio environment in space, maintaining useful transmit speeds can induce even higher power consumption, which further necessitates careful management of inter-satellite link usage.

\noindent\textbf{Parameter selection.}
For the inter-satellite link transmission speed, we select the one that enables low-power transmission below $0.1$ to minimize inter-satellite communication overhead and save satellite energy.
This leads to a rough speed of $2$ Mbps for S-Band and $50$ and $5$ Kbps for LoRa, respectively, while in both channels, a trade-off between speed and power is needed to minimize overall data transmission power consumption.
And we leave exploration for such trade-offs for future work.

\section{CPU Speed Function Fitting}
\label{appdx:cpu_speed_fitting}
\begin{figure}[ht]
\centering
\includegraphics[width=0.3\textwidth]{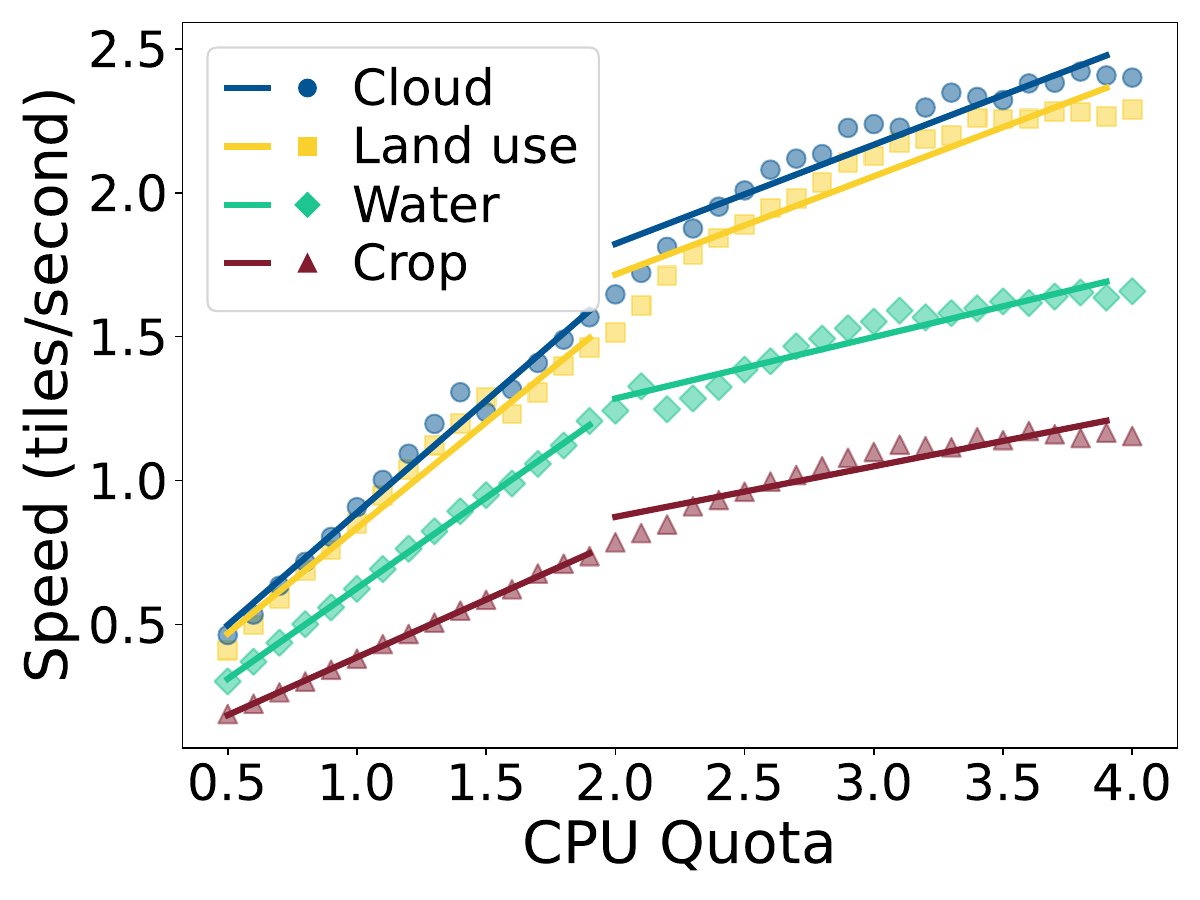}
\vspace{-1em}
\caption{Speed-CPU quota relation with piecewise linear fitting.}
\label{fig:fitting}
\end{figure}
\noindent
The two-piece piecewise linear fitting results for example analytics functions are presented in Fig.~\ref{fig:fitting}. 
The function parameters and coefficient of determination are shown in Table~\ref{tab:fitting_results}. 
The coefficients of determination generally exceed $0.9$ for profiled analytics functions, indicating an accurate model fit.

\begin{table}[]
    \centering
    \captionsetup{type=table}
    \begin{tabular}{|c|c|c|c|c|}
    \hline
    Function & Segment & Slope & Intercept & $R^2$ \\ \hline
    \multirow{2}{*}{Cloud} 
        & 0.5--2 & 0.7804 & 0.1073 & 0.9857 \\ \cline{2-5}
        & 2--4   & 0.3445 & 1.1331 & 0.9104 \\ \hline
    \multirow{2}{*}{Land use} 
        & 0.5--2 & 0.7338 & 0.1015 & 0.9805 \\ \cline{2-5}
        & 2--4   & 0.3414 & 1.0329 & 0.9020 \\ \hline
    \multirow{2}{*}{Object} 
        & 0.5--2 & 0.4012 & -0.0157 & 0.9994 \\ \cline{2-5}
        & 2--4   & 0.1758 & 0.5219  & 0.8692 \\ \hline
    \multirow{2}{*}{Water} 
        & 0.5--2 & 0.6300 & -0.0043 & 0.9990 \\ \cline{2-5}
        & 2--4   & 0.2136 & 0.8578  & 0.8995 \\ \hline
    \end{tabular}
    \caption{Piece-wise linear fitting parameters and corresponding $R^2$ scores across different analytics functions. Full function definitions are referred to Fig.~\ref{fig:profiling}.
    }
    \label{tab:fitting_results}
\end{table}

\section{Computing Workload Factors}
\label{appdx:real_graph_capacity}

\begin{algorithm}[!t]
\small
\SetAlgoNoEnd
\caption{\mbox{Compute workload factors of a workflow}}
\label{alg:real_graph_capacity}

\KwIn{Workflow graph $G_A = (\mathbf{M}, \mathbf{E})$, distribution ratios $\mathbf{D} = \{ \delta_{i,i'}\,|\, (m_i, m_{i'}) \in \mathbf{E}\}$}
\KwOut{Workload factors $\{ \rho_i \,|\, m_i \in \mathbf{M} \}$}

$\rho_i \leftarrow 0$ for $\forall m_i \in \mathbf{M}$\;

$Q \leftarrow \{ (m_i, 1.0) \,|\, m_i \text{ has in-degree } 0 \}$\;

\While{ $Q \ne \emptyset$}{ \label{alg:real_graph_capacity:begin_bfs}

    $m_i, \rho_i \leftarrow Q.pop()$

    \For{$m_{i'} \in downstream(m_i)$}{        
        $\rho_{i'} \leftarrow \rho_{i'} + \rho_i \cdot \delta_{i, i'}$\;
    }
}  \label{alg:real_graph_capacity:end_bfs}
    
\Return{$\{ \rho_i \,|\, m_i \in \mathbf{M} \}$}

\end{algorithm}

\noindent
Algorithm~\ref{alg:real_graph_capacity} specifies the algorithm for computing the workload factor of each analytics function in a workflow. %
The distribution ratios $\mathbf{D} = \{ \delta_{i,i'}\,|\, (m_i, m_{i'}) \in \mathbf{E}\}$ are independently profiled for each edge based on historical data, and once profiled can be used to flexibly estimate workload of the same function across different analytics workflows.
In Algorithm~\ref{alg:real_graph_capacity}, a BFS process calculates the average fraction of tiles into each analytics function instance given one source tile from the sensor.
For each analytics function, the BFS computes its workload factor as the sum of workload factors of each of its direct upstream functions, discounted by the distribution factor of each incoming edge.
The BFS has a time complexity of $O(|\mathbf{M}| + |\mathbf{E}|)$ and can be implemented efficiently.

\section{{\oursol} Planning and Control}
\label{appdx:implementation}
During planning, we solve Problem~\eqref{eq:objective} and execute Algorithm~\ref{alg:routing} to generate a set of pipelines for analytics function placement and workload routing. 
The planning results are then transmitted to satellites via control commands. 
In this section, we discuss the planning and control mechanisms of {\oursol}.

\subsection{On-ground Planning}
\label{appdx:implementation:planning}
\noindent\textbf{Planning frequency.}
In the planning phase, the analytics function deployment problem is solved and the routing algorithm runs on the ground, typically in a data center.
Planning is rerun whenever the constellation topology or workflow changes.
Since {\oursol} leverages a fully in-orbit analytics framework, the constellation topology remains relatively stable as it is not affected by frequent updates in ground-satellite connections.
Workflow updates can be scheduled daily or weekly for newly trained models~\cite{li2025adaorb} or new analytics pipelines.

\begin{figure}[ht]
\centering
\subfloat[Solving Program~\ref{eq:objective}.]{\includegraphics[width=0.232\textwidth]{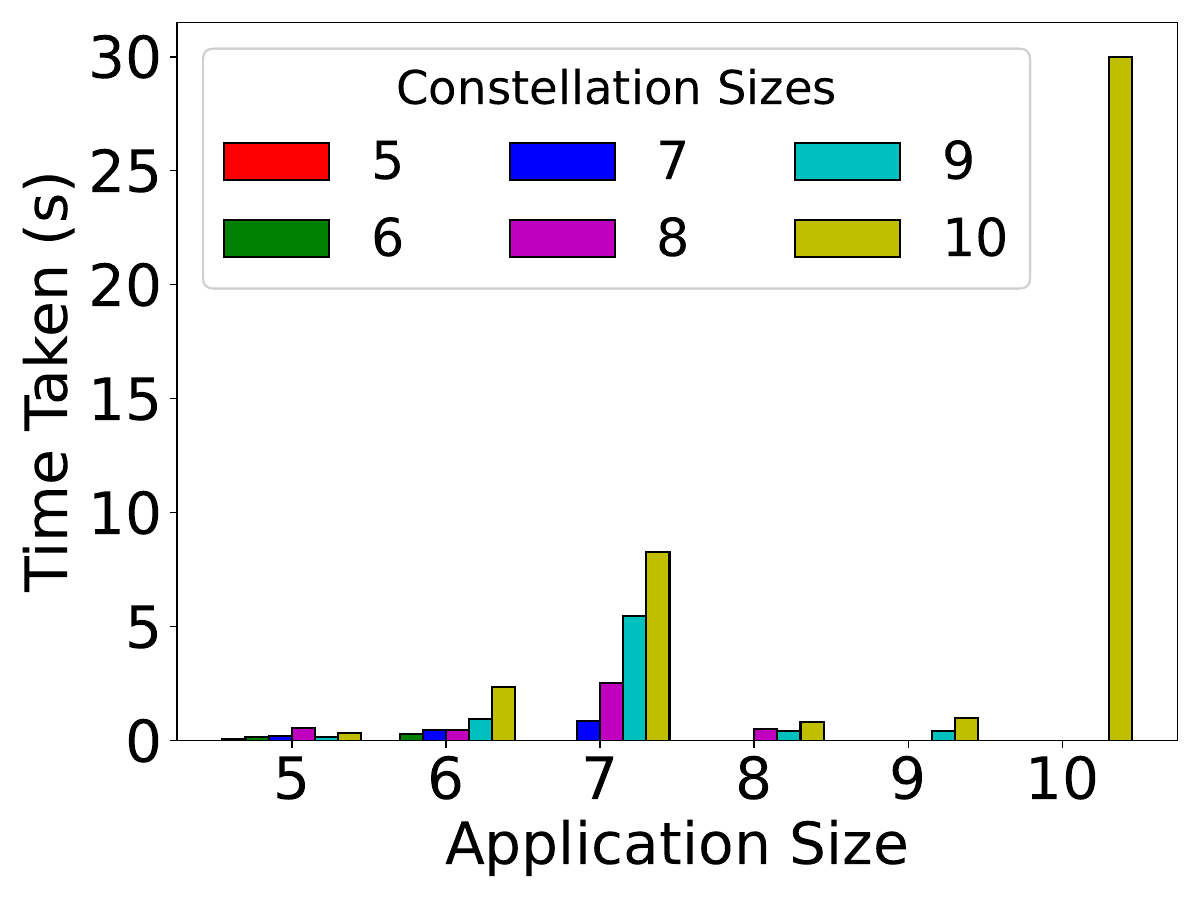}
\label{fig:planning_efficiency:placement}}
\hfil
\subfloat[Running Algorithm~\ref{alg:routing}.]{\includegraphics[width=0.232\textwidth]{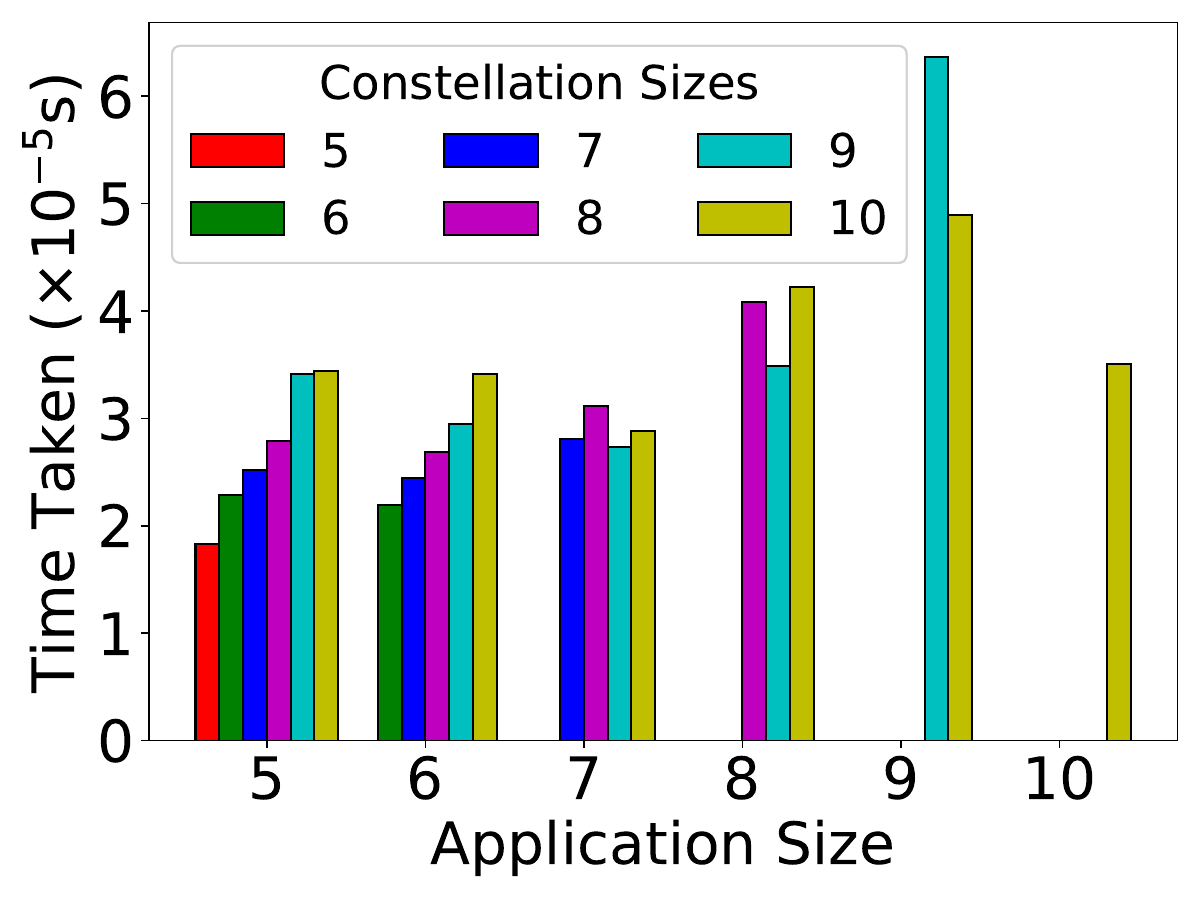}
\label{fig:planning_efficiency:routing}}
\hfil
\caption{The time required for solving Program~\ref{eq:objective} and executing Algorithm~\ref{alg:routing} for different constellation sizes (number of satellites in the constellation) and application sizes (number of analytics functions in the workflow). To ensure feasible solutions, we maintain fewer functions than satellites. The mixed integer program solves in under $30$ seconds for a $10$-satellite constellation, while the routing algorithm executes in less than one millisecond.}
\label{fig:planning_efficiency}
\end{figure}
\noindent\textbf{Planning efficiency.}
We conducted experiments to evaluate the efficiency of the mixed integer program~\eqref{eq:objective} and routing algorithm~\ref{alg:routing} across various satellite constellation sizes and analytics workflows.
We measured computation time for both program solving and algorithm execution. Our evaluation focused on constellations with satellite and analytics function counts ranging from $5$ to $10$.
We selected these sizes for two key reasons: larger constellations require more effort to maintain a leader-follower organization, and most existing workflows contain fewer than $10$ analytics functions~\cite{2021microservice}.
To ensure feasible solutions for Program~\ref{eq:objective} (which would otherwise cause early termination and prevent running Algorithm~\ref{alg:routing}), we maintained fewer analytics functions than satellites. 
All evaluations were performed on an off-the-shelf desktop with Intel 9900K CPU and 64GB memory.
The efficiency results for Program~\ref{eq:objective} and Algorithm~\ref{alg:routing} are presented in Fig.~\ref{fig:planning_efficiency}\subref{fig:planning_efficiency:placement} and Fig.~\ref{fig:planning_efficiency}\subref{fig:planning_efficiency:routing}, respectively.
As shown in Fig.~\ref{fig:planning_efficiency}\subref{fig:planning_efficiency:routing}, the routing algorithm executes in less than one millisecond across all test cases. 
Additionally, Fig.~\ref{fig:planning_efficiency}\subref{fig:planning_efficiency:placement} demonstrates that solving Program~\ref{eq:objective} for a 10-satellite 10-function constellation takes less than 30 seconds.
This overhead is negligible compared to the daily or weekly frequency of topology and workflow updates discussed previously.

\subsection{Constellation Control}
\label{appdx:implementation:constellation_control}
After obtaining the planning result, we need to propagate these results along with analytics functions to satellites for deployment and subsequent Earth observation analytics via constellation control channels.

\noindent\textbf{Constellation control implementation.}
Ground station control of LEO satellites primarily relies on the Telemetry, Tracking, and Command (TT\&C) system~\cite{yue2023low}. The primary protocol standards for command generation and uplink communication follow CCSDS recommendations—specifically the \textit{Telecommand (TC) Protocol} (CCSDS~232.0-B) and the \textit{Space Packet Protocol} (CCSDS~133.0-B)~\cite{2021tc}.
The \textit{S-band} is commonly used for its favorable propagation characteristics and allocation for space operations.
IEEE defines this band within the range from 2~GHz to 4~GHz, with typical frequency allocations of 2025 to 2110~MHz for uplink and 2200 to 2290~MHz for downlink~\cite{al2022survey}.
The timing of TT\&C constellation control commands depends critically on satellite pass windows—periods when a satellite is within visibility range of a ground station~\cite{mohamadhashim2023satellite, tao2023transmitting}.
For LEO satellite constellations, ground stations can predict visibility windows using satellite orbit (ephemeris) data, allowing them to organize, prioritize, and queue commands accordingly.
When a satellite enters a predicted visibility window, ground station antennas track it, establish an uplink, and transmit the queued control data.
Satellites then either execute received commands immediately or according to pre-scheduled onboard timing, and provide telemetry acknowledgments during the same or subsequent passes.

\end{document}